\documentclass[reprint,superscriptaddress,showpacs]{revtex4-1}
\usepackage{amsmath}
\usepackage{graphicx}
\usepackage{amssymb}

\begin{document}

\title{Entanglement and quantum discord in optically coupled coherent Ising machines}

\author{Yoshitaka Inui}
\email[]{yoshitaka.inui@ntt-research.com}

\affiliation{Physics and Informatics Laboratories, NTT Research Inc., 1950 University Ave., E. Palo Alto, CA 94303, USA}
\affiliation{National Institute of Informatics, Hitotsubashi 2-1-2, Chiyoda-ku, Tokyo 101-8430, Japan}

\author{Yoshihisa Yamamoto}

\affiliation{Physics and Informatics Laboratories, NTT Research Inc., 1950 University Ave., E. Palo Alto, CA 94303, USA}
\affiliation{E. L. Ginzton Laboratory, Stanford University, Stanford, CA 94305, USA}

\date{\today}

\begin{abstract}

We present analytical and numerical simulation results for 
squeezing, entanglement, and quantum discord in a dissipatively coupled coherent Ising machine (CIM). 
Both analytical solutions and numerical simulation results, which are obtained with positive-$P$, 
truncated-Wigner and truncated-Husimi representations for the density operator, 
predict the presence of entanglement and quantum discord, below and above the threshold of CIM. 
The entanglement criteria and quantum discord are evaluated as a function of  
the dissipative coupling strength relative to the background loss. 
For coupled two DOPOs, while entanglement disappears as the background loss exceeds the Ising coupling strength, 
the quantum discord remains finite even with a large linear loss. 
For one-dimensional lattice of DOPOs, 
while entanglement disappears for DOPO pair with large distance, quantum discord remains finite. 

\end{abstract}

\pacs{42.50.Ar, 64.90.+b, 03.67.Mn}

\maketitle

\section{Introduction}

A coherent Ising machine (CIM) with delay-line coupled degenerate optical parametric oscillators (DOPOs) 
has been theoretically studied \cite{Wang13,Takata15,Maruo16,Hamerly16} 
and experimentally demonstrated\cite{Marandi14,Takata16,Inagaki16}. 
In contrast to a CIM with measurement-feedback-coupled DOPOs\cite{McMahon16,Inagaki16b,Shoji17,Yamamura17}, 
a CIM with optically coupled DOPOs has been shown to satisfy 
an entanglement criterion \cite{Takata15,Maruo16}. 
The theoretical model of optically coupled CIM employs the linear Liouvillian coupling of squeezed states. 
The entanglement generation from squeezed states and linear Hamiltonian coupling 
has been considered in the transient case\cite{Furusawa98,Furusawa07,Paris99} and in the steady state case\cite{Olsen06}. 
The study of dissipative coupling induced entanglement generation started from atomic systems 
in the conditional case\cite{Cabrillo99,Plenio99} and in the steady state case\cite{Schneider02}.
The coherent Ising machine realizes dissipative coupling induced steady-state entanglement. 
The previous numerical results of CIM showed that it satisfies the sufficient condition for entanglement criterion\cite{Duan00}, 
not only below the threshold but also above it \cite{Takata15,Maruo16}. 
The degree of entanglement above the threshold depends on various system parameters: 
for some cases the entanglement criterion is violated before the threshold is reached\cite{Takata15}. 
The quantum discord \cite{Ollivier01,Henderson01} was also calculated \cite{Takata15}, 
but the difference in behavior with entanglement has not been clarified. 
To obtain the condition for entanglement and to understand a different behavior from the quantum discord in a CIM, 
a comprehensive theoretical decription which passes the numerical test 
for a wide range of system parameters is required. 

The previous theoretical investigation of CIM \cite{Takata15,Maruo16} 
depended on numerical simulation, particularly in the phase space method. 
A rigorous numerical simulation of cavity quantum electrodynamics (C-QED) system 
can be performed when the field density operator is expanded 
by Fock states with discrete spectra\cite{Dalibard92,Mu92,Molmer93}. 
The Fock state approach is effective for various nonlinear quantum optical effects \cite{Gisin92} 
if a system is in a small-photon-number regime. 
In CIMs, however, the system is in a large-photon-number regime where the Fock state approach is poorly suited. 
Open bosonic quantum systems, such as lasers and optical parametric oscillators (OPOs) 
are alternatively described by Heisenberg-Langevin equations in the Heisenberg picture \cite{Lax66} or 
by $c$-number SDEs in the Schr\"odinger picture \cite{Lax69}. 
For lasers with a vastly increasing number of photons, the field density operator 
can be expanded by using diagonal coherent state expansion \cite{Glauber63,Lax67}, 
and the Fokker-Planck equation for the Glauber's $P$ function is derived. 
By subsequently using the Ito rule, the $c$-number stochastic differential equation (SDE) can be derived. 
For a DOPO, however, the Fokker-Planck equation of the $P$ function has a negative diffusion coefficient, and this approach fails. 

A CIM with dissipatively coupled DOPOs was first studied by using $c$-number Langevin equations \cite{Wang13}, 
which constitute a $c$-number counterpart to the $q$-number Heisenberg-Langevin equations\cite{Lax66}. 
Later, an equivalence between the $c$-number Heisenberg-Langevin equations and 
SDEs in Wigner representation was established with the truncation of Fokker-Planck equation in the Wigner representation\cite{Maruo16}. 
The phase space treatments of DOPOs \cite{McNeil78,Drummond79,Drummond81,Milburn81,Wolinsky88,Drummond89,Kinsler91} 
were, on the other hand, performed with positive-$P$ \cite{Drummond80,Drummond81b} or complex-$P$\cite{Drummond81b} representations. 
The SDEs in the positive-$P$ representation have been introduced into the numerical simulation of a CIM \cite{Takata15,Maruo16}. 
These phase-space methods overcome the difficulties in Glauber's diagonal $P$ representation and can deal with 
a varying photon number over many orders of magnitude. 
Various quantum features can be computed from such SDEs if a sufficient number of trajectories are ensemble averaged. 
Positive-$P$ and truncated Wigner approaches produce similar values for entanglement criterion\cite{Maruo16}, 
if DOPOs are operating in a so-called weak noise limit\cite{Wolinsky88}. 

On the analytical side, quantum statistical properties of a single DOPO at a steady state have been 
rigorously obtained by the integration of the Fokker-Planck equation in the complex-$P$ representation\cite{Drummond81,Milburn81}. 
In coupled nonlinear quantum optical systems, some recent theoretical studies have used the mean field approximation 
for the coupling part between subsystems\cite{Sheshadri93,Boite13,Savona17}. 
Such treatment, motivated by condensed matter theory\cite{Sheshadri93}, is useful in searching for a macroscopic or global order, 
but neglects quantum correlation, including entanglement, between subsystems. 
To elucidate the quantum correlation among constituent DOPOs in a CIM, 
we consider a weak noise limit \cite{Wolinsky88} of a DOPO, 
instead of the general solution of a DOPO \cite{Drummond81,Milburn81}. 
In such a weak noise case, we can analytically calculate entanglement and quantum discord. 
To check the validity of analytical results, we compute the squeezing, entanglement and quantum discord 
numerically by using the Positive-$P$, truncated-Wigner and truncated-Husimi SDEs. 
The numerical simulation results agree completely with independently derived analytical solutions. 

The paper is organized as follows. 
In Sec.II, the quantum master equation for a single DOPO is presented. 
In Sec.III, we introduce the theoretical model of two dissipatively coupled DOPOs 
and present the analytical and numerical results on the degrees of squeezing/anti-squeezing, entanglement, and quantum discord. 
Analytical results are compared with numerical results.  
Section IV examines a one-dimensional (1D) lattice of DOPOs with only nearest-neighbor coupling. 
Section V summarizes the main results. 
Appendix A summarizes the numerical simulation methods based on the positive-$P$, truncated-Wigner and truncated-Husimi SDEs for a single DOPO. 
Appendix B summarizes the analytical method for a single DOPO. 
Appendix C shows the traveling wave model for 1D nearest-neighbor-coupled DOPOs. 

\section{Master equation of a DOPO}

In this section, we present the theoretical model for a single DOPO. 
The system of a single DOPO consists of 
the pump-mode operator $\hat{a}_p$ and the signal-mode operator $\hat{a}_s$. 
The quantum master equation of the system is represented as follows: 
\begin{eqnarray}
\label{QME0}
\frac{\partial \hat{\rho}}{\partial t}=\mathcal{L}\hat{\rho}=\varepsilon[\hat{a}_p^{\dagger}-\hat{a}_p,\hat{\rho}]+\frac{\kappa}{2}[\hat{a}_s^{\dagger 2}\hat{a}_p-\hat{a}_p^{\dagger}\hat{a}_s^2,\hat{\rho}]\nonumber \\
+(\gamma_p[\hat{a}_p,\hat{\rho}\hat{a}_p^{\dagger}]+\gamma_s[\hat{a}_s,\hat{\rho}\hat{a}_s^{\dagger}]+{\rm h.c.}). 
\end{eqnarray}
Here, $[\hat{A},\hat{B}]=\hat{A}\hat{B}-\hat{B}\hat{A}$. This equation has Hamiltonian and Liouvillian parts. 
The Hamiltonian part has two terms $\hat{H}_1+\hat{H}_2$, where 
$\hat{H}_1=i\hbar \varepsilon(\hat{a}_p^{\dagger}-\hat{a}_p)$ is the coherent excitation of the pump mode by an external injection field $\varepsilon$ 
and $\hat{H}_2=i\hbar \frac{\kappa}{2}(\hat{a}_s^{\dagger 2}\hat{a}_p-\hat{a}_p^{\dagger}\hat{a}_s^2)$ 
is the nonlinear parametric coupling between the 
signal and pump modes via the second-order nonlinear coupling constant $\kappa$. 
The Liouvillian parts consist of the dissipation of the pump mode, represented by the half width at half maximum (HWHM) $\gamma_p$, 
and that of the signal mode, represented by $\gamma_s$.  
In the Heisenberg picture, the corresponding Heisenberg-Langevin equations \cite{Lax66} are as follows\cite{Wang13}. 
\begin{eqnarray}
\frac{d\hat{a}_p}{dt} &=& -\gamma_p\hat{a}_p+\varepsilon-\frac{\kappa}{2}\hat{a}_s^{2}+\sqrt{\gamma_p}\hat{\xi}_1 , \\
\frac{d\hat{a}_s}{dt} &=& -\gamma_s\hat{a}_s+\kappa\hat{a}_s^{\dagger}\hat{a}_p+\sqrt{\gamma_s}\hat{\xi}_2 , 
\end{eqnarray}
where $\langle \hat{\xi}_i^{\dagger}(t)\hat{\xi}_j(t')\rangle=0$, and $\langle \hat{\xi}_i(t)\hat{\xi}_j^{\dagger}(t')\rangle=2\delta_{ij}\delta(t-t')$. 
The injected pump field at the oscillation threshold is $\varepsilon_{thr}=\gamma_p\gamma_s/\kappa$. 
Assuming that $\gamma_p$ is sufficiently large compared to $\gamma_s$, 
the pump mode can be eliminated via 
\begin{equation}
\label{apae}
\hat{a}_p=\frac{\varepsilon}{\gamma_p}-\frac{\kappa}{2\gamma_p}\hat{a}_s^{2}+\sqrt{\frac{1}{\gamma_p}}\hat{\xi}_1. 
\end{equation}
After eliminating the pump mode adiabatically by substituting Eq.(\ref{apae}) 
into $\hat{H}_2$ in the quantum master equation (\ref{QME0}) and averaging over the pump mode and the noise operator $\hat{\xi}_1$, 
we can obtain the quantum master equation for only the signal mode (below, we omit the signal mode subscript $s$ ) \cite{Drummond81}: 
\begin{equation}
\label{QME}
\mathcal{L}_{DOPO}\hat{\rho}=\frac{S}{2}[\hat{a}^{\dagger 2}-\hat{a}^2,\hat{\rho}]+\Bigl(\gamma_s[\hat{a},\hat{\rho}\hat{a}^{\dagger}]+\frac{B}{2}[\hat{a}^2,\hat{\rho}\hat{a}^{\dagger 2}]+{\rm h.c.}\Bigr) . 
\end{equation}
Here, the parameter $S=\gamma_s\varepsilon/\varepsilon_{thr}$ represents the squeezing/anti-squeezing rate 
due to the parametric interaction between the pump mode and the signal mode, and 
$B=\kappa^2/(2\gamma_p)$ represents the degenerate two photon absorption that describes the saturation of the parametric gain. 
This master equation is equivalent to the following Heisenberg-Langevin equation\cite{Wang13}: 
\begin{equation}
\label{HL}
\frac{d\hat{a}}{dt}=-\gamma_s\hat{a}+S\hat{a}^{\dagger}-B\hat{a}^{\dagger}\hat{a}^2+\sqrt{2B}\hat{a}^{\dagger}\hat{\xi}_1+\sqrt{\gamma_s}\hat{\xi}_2 . 
\end{equation}
The $c$-number counterpart of this equation is equivalent to the SDE in the truncated Wigner representation\cite{Maruo16}. 
The three phase space methods (positive-$P$, truncated Wigner, and truncated Husimi) 
for dealing with the quantum master equation [Eq.(\ref{QME})] are presented in Appendix A.  

\section{Two coupled DOPOs}

\subsection{Model}

\begin{figure}
\begin{center}
\includegraphics[width=7cm]{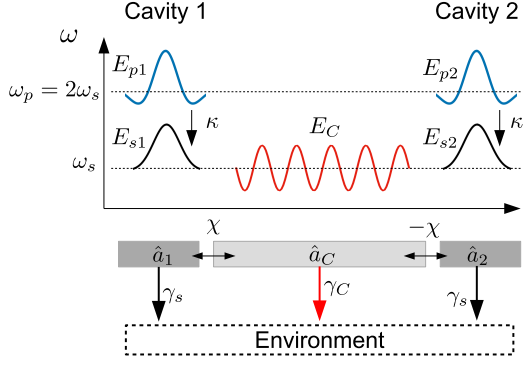}
\caption{Standing-wave model for ferromagnetically coupled DOPOs. 
The two cavities support pump and signal modes, and the two signal modes $\hat{a}_r(r=1,2)$ 
are coupled via the coupling mode $\hat{a}_C$. 
The eigenfunction of $\hat{a}_C$ has a node at the center of the cavity. 
When the two pump modes are excited with the same phase, 
the two signel modes $\hat{a}_1$ and $\hat{a}_2$ can have the same or opposite phases. 
When $\hat{a}_2$ has the same phase as $\hat{a}_1$, the coupling mode $\hat{a}_C$ is not excited, 
so that the dissipation by a decay rate $\gamma_C$ disappears. }
\label{model1}
\end{center}
\end{figure}

We now consider a simplest CIM consisting of two DOPOs with ferromagnetic dissipative coupling, as shown in Fig.\ref{model1}. 
For the two signal modes represented by $\hat{a}_1$ and $\hat{a}_2$, 
the Liouvillian of an entire system is the sum of a single DOPO part and coupling part: 
$\frac{\partial \hat{\rho}}{\partial t}=\sum_r \mathcal{L}_{DOPO}^{(r)}\hat{\rho}+\mathcal{L}_{C}\hat{\rho}$. 
Here, $\mathcal{L}_{DOPO}^{(r)}$ operates on only $r$-th DOPO and $\mathcal{L}_C\hat{\rho}$ operates on both. 
We consider the following Liouvillian which represents the dissipative ferromagnetic coupling\cite{Takata15}: 
\begin{equation}
\label{lc}
\mathcal{L}_{C}\hat{\rho}=J[\hat{a}_{1}-\hat{a}_{2},\hat{\rho}(\hat{a}_{1}^{\dagger}-\hat{a}_{2}^{\dagger})]+{\rm h.c.} . 
\end{equation}
This Liouvillian is derived when the two signal modes $\hat{a}_1$ and $\hat{a}_2$ with eigenfunctions $E_{s1}$ and $E_{s2}$, respectively, are 
coupled via the standing-wave cavity mode $\hat{a}_C$ with eigenfunction $E_C$, 
which has a node at the center of the cavity and has the oppositely displaced anti-nodes at the interface of $\hat{a}_1$ and $\hat{a}_2$: 
\begin{equation}
\hat{H}_C=\hbar \chi(\hat{a}_1^{\dagger}\hat{a}_C+\hat{a}_C^{\dagger}\hat{a}_1)-\hbar \chi(\hat{a}_2^{\dagger}\hat{a}_C+\hat{a}_C^{\dagger}\hat{a}_2). 
\end{equation}
Here, the coupling mode $\hat{a}_C$ has a dissipation rate (HWHM) $\gamma_C$. 
When the two signal modes $\hat{a}_{1}$ and $\hat{a}_{2}$ are excited with the same phase (ferromagnetic order), 
the dissipation induced by $\gamma_C$ does not occur. 
When $\gamma_C$ is large, mode $\hat{a}_C$ can be adiabatically eliminated via the Lindblad procedure, 
and then equation (\ref{lc}) is obtained with $J\sim \chi^2/\gamma_C$. 

The positive-$P$ SDE for two coupled DOPOs is obtained, 
as the extension of that for a single DOPO, shown in Appendix A, 
as 
\begin{eqnarray}
\frac{d}{dt}
\begin{bmatrix}
\alpha_1 \\ \alpha_1^{\dagger}
\end{bmatrix}
&=& -
\begin{bmatrix} 
\gamma_s+J+B\alpha_1^{\dagger}\alpha_1 & -S \\
 -S & \gamma_s+J+B\alpha_1^{\dagger}\alpha_1
\end{bmatrix}
\begin{bmatrix}
\alpha_1 \\ \alpha_1^{\dagger}
\end{bmatrix}
\nonumber \\
&+& J
\begin{bmatrix}
\alpha_2 \\ \alpha_2^{\dagger}
\end{bmatrix}
+
\begin{bmatrix}
\sqrt{S-B\alpha_1^2}\xi_{R1} \\
\sqrt{S-B\alpha_1^{\dagger 2}}\xi_{R1}^{\dagger}
\end{bmatrix},
\end{eqnarray}
\begin{eqnarray}
\frac{d}{dt}
\begin{bmatrix}
\alpha_2 \\ \alpha_2^{\dagger}
\end{bmatrix}
&=&-
\begin{bmatrix} 
\gamma_s+J+B\alpha_2^{\dagger}\alpha_2 & -S \\
 -S & \gamma_s+J+B\alpha_2^{\dagger}\alpha_2
\end{bmatrix}
\begin{bmatrix}
\alpha_2 \\ \alpha_2^{\dagger}
\end{bmatrix}
\nonumber \\
&+& J
\begin{bmatrix}
\alpha_1 \\ \alpha_1^{\dagger}
\end{bmatrix}
+
\begin{bmatrix}
\sqrt{S-B\alpha_2^2}\xi_{R2} \\
\sqrt{S-B\alpha_2^{\dagger 2}}\xi_{R2}^{\dagger}
\end{bmatrix}.
\end{eqnarray}
Here $\alpha_r$ and $\alpha_r^{\dagger} (r=1,2)$ are independent $c$-numbers satisfying 
$\langle \alpha_r \rangle=\langle \alpha_r^{\dagger}\rangle^*$ and 
$\xi_{Rr}$ and $\xi_{Rr}^{\dagger}$ are independent real number random variables satisfying 
$\langle \xi_{Rr}(t)\xi_{Rr'}(t')\rangle=\delta_{rr'}\delta(t-t')$ 
and $\langle \xi_{Rr}^{\dagger}(t)\xi_{Rr'}^{\dagger}(t')\rangle=\delta_{rr'}\delta(t-t')$. 
When the system has the ferromagnetic order $\alpha_1\sim\alpha_2$, each cavity has no additional loss due to coupling. 
Therefore, the oscillation occurs with the same excitation as for a single DOPO i.e., 
at normalized pump rate $p=S/\gamma_s=1$. 
The dissipative coupling Liouvillian (\ref{lc}) does not introduce new noise terms into the positive-$P$ SDE \cite{Takata15}. 
In the Wigner and Husimi SDEs, however, the dissipative coupling introduces new noise terms. 
When the coupling parts of the SDE are written as $\frac{d\alpha_1}{dt}|_C$ and $\frac{d\alpha_2}{dt}|_C$, they are represented as: 
\begin{eqnarray}
\frac{d\alpha_1}{dt}|_C &=& -J\alpha_1+J\alpha_2+\sqrt{AJ}\xi_C , \\
\frac{d\alpha_2}{dt}|_C &=& -J\alpha_2+J\alpha_1-\sqrt{AJ}\xi_C . 
\end{eqnarray}
Here, $A=1$ for the Husimi representation and $A=\frac{1}{2}$ for the Wigner representation. 
The complex noise source $\xi_C$, 
satisfying $\langle \xi^*_C(t)\xi_C(t')\rangle=2\delta(t-t') $, 
is common for $\alpha_1$ and $\alpha_2$, 
since it comes from the vacuum noise of the coupling mode $\hat{a}_C$. 

\subsection{Covariance matrix}

We consider the steady-state covariance matrix for two coupled DOPOs 
by applying the fluctuation analysis in the positive-$P$ representation, 
that for a single DOPO is presented in Appendix B. 
If two modes have the same excitations, then they have symmetry under the exchange of modes. 
We assume that the fluctuation products of the positive-$P$ amplitudes 
satisfy $\langle \Delta \alpha_{1}^2\rangle=\langle \Delta \alpha_{2}^2\rangle$ and 
$\langle \Delta \alpha_{1}^{\dagger}\Delta \alpha_{1}\rangle=\langle \Delta \alpha_{2}^{\dagger}\Delta \alpha_{2} \rangle$. 
The covariance matrix $\sigma$ is defined for the vectors $\overrightarrow{\hat{R}}=\sqrt{2}[\hat{X}_1,\hat{P}_1,\hat{X}_2,\hat{P}_2]$, 
\begin{equation}
\label{cov}
\sigma_{ij}=\frac{1}{2}\langle [\hat{R}_i, \hat{R}_j]_+\rangle-\langle \hat{R}_i\rangle \langle \hat{R}_j\rangle. 
\end{equation}
Here, $[\hat{A},\hat{B}]_+=\hat{A}\hat{B}+\hat{B}\hat{A}$.  
For two DOPOs with dissipative coupling, this matrix can be written as follows: 
\begin{equation}
\label{cov2}
\sigma=
\begin{bmatrix}
\underline{\alpha} & \underline{\gamma} \\
\underline{\gamma} & \underline{\alpha}
\end{bmatrix}. 
\end{equation}
Here, $\underline{\alpha}={\rm diag}(a_1,a_2)$ and $\underline{\gamma}={\rm diag}(c_1,c_2)$. 
$a_1=2\langle \Delta \hat{X}_1^2\rangle=2\langle \Delta \hat{X}_2^2\rangle$, and 
$a_2=2\langle \Delta \hat{P}_1^2\rangle=2\langle \Delta \hat{P}_2^2\rangle$ 
represent the single-mode variances of $\hat{X}$ and $\hat{P}$, respectively. 
On the other hand, 
$c_1=2\langle \Delta \hat{X}_1\Delta \hat{X}_2\rangle$, and 
$c_2=2\langle \Delta \hat{P}_1\Delta \hat{P}_2\rangle$ 
represent the intermode correlation of $\hat{X}$ and $\hat{P}$, respectively. 

Below the oscillation threshold ($p<1$), where $\langle \alpha_{1}\rangle=\langle \alpha_{2}\rangle=0$, 
the steady-state amplitude fluctuations satisfy the following: 
\begin{equation}
\begin{bmatrix}
1+j & -p & -j & 0\\
-p & 1+j & 0 & -j\\
-j & 0 & 1+j & -p\\
0 & -j & -p & 1+j
\end{bmatrix}
\begin{bmatrix}
\langle \Delta \alpha_1^2\rangle \\
\langle \Delta \alpha_1^{\dagger} \Delta \alpha_1 \rangle \\
\langle \Delta \alpha_1 \Delta \alpha_2 \rangle \\
\langle \Delta \alpha_1^{\dagger} \Delta \alpha_2 \rangle 
\end{bmatrix}
=\frac{p}{2}
\begin{bmatrix}
1\\ 0\\0 \\0
\end{bmatrix}. 
\end{equation}
Here, $j:=J/\gamma_s$ is the normalized coupling coefficient. 
$\langle \Delta \alpha_1^2\rangle$ and $\langle \Delta \alpha_1^{\dagger} \Delta \alpha_1 \rangle $ 
represent the averaged fluctuation products for the same DOPO, 
while $\langle \Delta \alpha_1 \Delta \alpha_2 \rangle$ and $\langle \Delta \alpha^{\dagger}_1 \Delta \alpha_2 \rangle$ 
represent the averaged fluctuation products for different DOPOs. 
For the system of two DOPOs below the threshold, 
all the nonzero components of the covariance matrix are obtained as follows. 
\begin{equation}
\label{alphab}
\underline{\alpha}={\rm diag}\Bigl[1+\frac{p(1-p+j)}{(1-p)(1-p+2j)},1-\frac{p(1+p+j)}{(1+p)(1+p+2j)}\Bigr],
\end{equation}
\begin{equation}
\label{gammab}
\underline{\gamma}={\rm diag}\Bigl[\frac{pj}{(1-p)(1-p+2j)},-\frac{pj}{(1+p)(1+p+2j)}\Bigr].
\end{equation}

Above the threshold, the mean amplitudes are given by 
$\langle \alpha_1\rangle=\langle \alpha_2\rangle=\sqrt{\frac{\gamma_s}{B}(p-1)}$. 
We consider the fluctuation around this mean value using the linearization technique \cite{Chaturvedi77}. 
This procedure in a single DOPO is shown in Appendix B. 
Under a steady-state condition, the fluctuation amplitude products are represented as follows: 
\begin{widetext}
\begin{equation}
\begin{bmatrix}
2p-1+j & -1 & -j & 0\\
-1 & 2p-1+j & 0 & -j\\
-j & 0 & 2p-1+j & -1\\
0 & -j & -1 & 2p-1+j
\end{bmatrix}
\begin{bmatrix}
\langle \Delta \alpha_1^2\rangle \\
\langle \Delta \alpha_1^{\dagger} \Delta \alpha_1 \rangle \\
\langle \Delta \alpha_1 \Delta \alpha_2 \rangle \\
\langle \Delta \alpha_1^{\dagger} \Delta \alpha_2 \rangle 
\end{bmatrix}
=\frac{1}{2}
\begin{bmatrix}
1\\ 0\\0 \\0
\end{bmatrix}. 
\end{equation}
\end{widetext}
These amplitude fluctuation products produce the following nonzero components of covariance matrix: 
\begin{equation}
\label{alphaa}
\underline{\alpha}={\rm diag}\Bigl[1+\frac{2p-2+j}{4(p-1)(p-1+j)},1-\frac{2p+j}{4p(p+j)}\Bigr], 
\end{equation}
\begin{equation}
\label{gammaa}
\underline{\gamma}={\rm diag}\Bigl[\frac{j}{4(p-1)(p-1+j)},-\frac{j}{4p(p+j)}\Bigr].
\end{equation}

\subsection{Sufficient condition for entanglement}

The sufficient criterion for entanglement is given in Ref.\cite{Duan00}, using  
\begin{equation}
\mathfrak{D}=\langle \Delta \hat{u}^2\rangle+\langle \Delta \hat{v}^2\rangle, 
\end{equation}
where $\hat{u}=\hat{X}_1-\hat{X}_2$, $\hat{v}=\hat{P}_1+\hat{P}_2$, 
$\hat{X}_r=\frac{\hat{a}_r+\hat{a}_r^{\dagger}}{\sqrt{2}}(r=1,2)$ and $\hat{P}_r=\frac{\hat{a}_r-\hat{a}_r^{\dagger}}{\sqrt{2}i}(r=1,2)$. 
If $\mathfrak{D}/2<1$, then the system's density matrix is not separated into 
the product states of the individual DOPOs : $\hat{\rho}\ne \hat{\rho}_1\otimes \hat{\rho}_2$. 
This criterion can be calculated using the components of covariance matrix as 
\begin{equation}
\frac{\mathfrak{D}}{2}= \frac{(a_1-c_1)+(a_2+c_2)}{2}. 
\end{equation}
From Eqs.(\ref{alphab})(\ref{gammab}), below threshold $\mathfrak{D}/2$ is represented as follows: 
\begin{equation}
\label{duanb}
\frac{\mathfrak{D}}{2}=1-\frac{p(j-p)}{(1+p)(1-p+2j)}. 
\end{equation}
Then the sufficient condition for entanglement is satisfied if $j>p$ is satisfied. 
This means that when $j<1$, the entanglement criterion is violated before the threshold is reached as seen in Ref.\cite{Takata15}. 
Above the threshold, the entanglement criterion is represented as: 
\begin{equation}
\label{duana}
\frac{\mathfrak{D}}{2}=1-\frac{j-1}{4p(p-1+j)}. 
\end{equation}
We can thus see that the entanglement criterion is satisfied even above the threshold, 
when the dissipative coupling rate is larger than a linear loss rate $j=J/\gamma_s>1$. 

\subsection{Necessary and sufficient entanglement criterion}

We now consider the necessary and sufficient 
entanglement criterion following Simon's theory\cite{Simon00,Vidal02,Serafini03,Adesso04}. 
For symmetric two boson gaussian system, this is easily 
shown to be equivalent to "Theorem 2" in Ref.\cite{Duan00}. 
In the description in Ref.\cite{Adesso04},
the necessary and sufficinet entanglement criterion is 
described by the symplectic eigenvalues of 
covariance matrix $\sigma_{ij}$. 
The covariance matrix can be diagonalized with a symplectic transformation $\Sigma$ as 
$\Sigma \sigma \Sigma^T={\rm diag} (\nu_+,\nu_+,\nu_-,\nu_-)$, where $\nu_+>\nu_-$.
The symplectic eigenvalues are related to the uncertainty relation, and a physical state must satisfy $\nu_- \ge 1$ \cite{Adesso04}. 
These eigenvalues $\nu_+$ and $\nu_-$ are obtained as the eigenvalues of $i \Omega \sigma$ where 
$\Omega=\begin{bmatrix} 0 & 1\\-1 & 0\end{bmatrix}\oplus \begin{bmatrix} 0 & 1\\-1 & 0\end{bmatrix}$.  
Simon's entanglement criterion is defined on the partially transposed covariance matrix 
$\bar{\sigma}=\Lambda \sigma \Lambda^T$, where $\Lambda={\rm diag}(1,1,1,-1)$. 
The symplectic eigenvalues of $\bar{\sigma}$ are eigenvalues of $i \Omega \bar{\sigma}$ and 
satisfy $\Sigma '\bar{\sigma} \Sigma'^T={\rm diag} (\tilde{\nu}_+,\tilde{\nu}_+,\tilde{\nu}_-,\tilde{\nu}_-)$ 
($\tilde{\nu}_+ > \tilde{\nu}_-$), for a symplectic transformation $\Sigma '$. 
The separable state must satisfy $\tilde{\nu}_- \ge 1$\cite{Adesso04}. 

The above procedures produce the symplectic eigenvalues of $\sigma$. 
The two independent symplectic eigenvalues of $\sigma$ are calculated as eigenvalues $\lambda$ of $i\Omega \sigma $, 
and are reduced to $\lambda^2=(a_1 + c_1)(a_2 + c_2), (a_1 - c_1)(a_2 - c_2)$. 
Substituting Eqs.(\ref{alphab})(\ref{gammab}) into $\lambda^2$, we obtain 
$\nu_+^2=1+\frac{p^2}{1-p^2}$ and $\nu_-^2=1+\frac{p^2}{(1+p+2j)(1-p+2j)}$. 
To obtain necessary and sufficient entanglement criterion, 
two independent symplectic eigenvalues of $\bar{\sigma}$ are calculated 
as eigenvalues $\lambda'$ of $i\Omega \bar{\sigma}$, which are reduced to 
$\lambda'^2=(a_1+c_1)(a_2-c_2), (a_1-c_1)(a_2+c_2)$. 
The smallest symplectic eigenvalue of $\bar{\sigma}$ is 
\begin{equation}
\label{simonb}
\tilde{\nu}_-^2=1-\frac{p(2j-p)}{(1+p)(1-p+2j)}. 
\end{equation}
When $\tilde{\nu}_-<1 $, the two coupled DOPOs are inseparable. 
This entanglement condition for below the threshold is satisfied when $j>p/2$. 

We next consider the above-threshold characteristics described by 
the covariance matrix $\sigma$. 
The symplectic eigenvalues of 
$\sigma$ are $\nu_+^2=1+\frac{1}{4p(p-1)}$ and $\nu_-^2=1+\frac{1}{4(p+j)(p-1+j)}$. 
The smallest symplectic eigenvalue of $\bar{\sigma}$ is represented as 
\begin{equation}
\label{simona}
\tilde{\nu}_-^2=1-\frac{2j-1}{4p(p-1+j)}. 
\end{equation}
Therefore, the necessary and sufficient entanglement criterion for above the threshold is satisfied for $j>1/2$. 
The required $j$ to satisfy the entanglement criterion is thus smaller than 
that for sufficient criterion obtained in the previous section ($j>1$). 
The symplectic eigenvalue of partially transposed covariance matrix $\bar{\sigma}$ 
represents the degree of entanglement and the minimum value of $\tilde{\nu}_-$ is $\tilde{\nu}_-\rightarrow \frac{1}{\sqrt{2}}$ 
when $p=1$ and $j\rightarrow \infty$. This is equivalent to that obtained for one-dimensional model in Ref.\cite{Yanagimoto19}. 

\subsection{Quantum discord}

Next, we consider the quantum discord of two dissipatively coupled DOPOs. 
When the total system density operator $\hat{\rho}$ 
is composed of the first DOPO with $\hat{\rho}_1$ and the second DOPO with $\hat{\rho}_2$, 
the von Neumann entropy of the total system is defined by $S(\hat{\rho})=-{\rm Tr} \hat{\rho} \log \hat{\rho}$. 
The total entropy also seems to be written as the summation $S(\hat{\rho}_1)+S(\hat{\rho}_{2|1})$, where
$S(\hat{\rho}_1)$ is the reduced entropy of the first DOPO, 
while $S(\hat{\rho}_{2|1})$ is the conditional (or residual) entropy 
of the second DOPO which is given by the subtraction of mutual information from $S(\hat{\rho}_2)$. 
In classical information theory, these two formula for the total entropy are identical, 
but when the systems have quantumness they generally differ\cite{Ollivier01,Henderson01}.
The quantum discord is defined as the difference between these two expressions for entropies. 
Non-zero quantum discord is identified as a metric 
for quantifying the quantum correlation of mixed states 
and as a useful quantum resource in a specific quantum computational model \cite{Datta07,Lanyon08}. 

For a system of two DOPOs, using the theory of two-mode gaussian states \cite{Giorda10,Adesso10}, 
quantum discord for the covariance matrix $\sigma$ is written as follows: 
\begin{equation}
\label{qdc}
\mathcal{D}(\sigma)=f(\sqrt{a_1 a_2})+\inf_{\underline{\sigma_0}} f(\sqrt{\det \underline{\varepsilon}})-f(\nu_{+})-f(\nu_{-}). 
\end{equation}
Here $f(x)=\frac{x+1}{2}\log \frac{x+1}{2}-\frac{x-1}{2}\log\frac{x-1}{2}$,  
and $\underline{\varepsilon}=\underline{\alpha}-\underline{\gamma}(\underline{\sigma_0}+\underline{\alpha})^{-1}\underline{\gamma}^T$
where $\underline{\sigma_0}$ relates to the measurement which removes the mutual information\cite{Eisert02,Giedke02}. 
The first two terms on the right hand side of Eq.(\ref{qdc}) represent the entropies $S(\hat{\rho}_1)$ and $S(\hat{\rho}_{2|1})$. 
The last two terms represent the entropy $S(\hat{\rho})$ consisting of two DOPOs\cite{Holevo01}. 
In the calculation of $S(\hat{\rho}_{2|1})$, we consider $\theta=0$ case of Ref.\cite{Adesso10} 
where $\underline{\sigma_0}={\rm diag}(\lambda,\lambda^{-1})$($\lambda>0$). 
When $(a_2 c_1^2-a_1c_2^2(a_1^2-c_1^2))(a_2 c_1^2(a_2^2-c_2^2)-a_1 c_2^2)<0$, 
$\frac{d\det \underline{\varepsilon}(\lambda)}{d\lambda}=0$ can be obtained for positive $\lambda$. 
When $(a_2 c_1^2-a_1c_2^2(a_1^2-c_1^2))(a_2 c_1^2(a_2^2-c_2^2)-a_1 c_2^2)\ge 0$, 
$\frac{d\det \underline{\varepsilon}(\lambda)}{d\lambda}=0$ is satisfied for only negative $\lambda$, 
and the minimum value of the determinant of the matrix $\underline{\varepsilon}$ is obtained at $\lambda \rightarrow 0$ \cite{Adesso10} as 
\begin{equation}
\label{infs}
\inf_{\underline{\sigma_0}} \det \underline{\varepsilon}=\frac{a_2}{a_1}(a_1^2-c_1^2). 
\end{equation} 
In our two-DOPO system, the residual entropy is minimized for $\lambda\rightarrow 0$. 
This is shown from the analytical result of 
$D:=a_2 c_1^2(a_2^2-c_2^2)-a_1 c_2^2 (a_1^2-c_1^2)$, 
which is smaller than both 
$a_2 c_1^2-a_1 c_2^2 (a_1^2-c_1^2)$, 
and $a_2 c_1^2(a_2^2-c_2^2)-a_1 c_2^2$, because of $a_1^2-c_1^2>1$ and $a_2^2-c_2^2<1$.  
Both $D=\frac{2p^3j^2(1+j)(1+2j)}{(1+p)^2(1-p)^2(1-p+2j)^2(1+p+2j)^2}$ below the threshold and 
$D=\frac{j^2(2p-1)(2p-1+j)(2p-1+2j)}{128p^2(p-1)^2(p+j-1)^2(p+j)^2}$ above the threshold have positive values. 
Therefore $(a_2 c_1^2-a_1c_2^2(a_1^2-c_1^2))(a_2 c_1^2(a_2^2-c_2^2)-a_1 c_2^2)\ge 0$ is always satisfied.  
The quantum discord is always nonzero in an optically coupled CIM, 
no matter how large the dissipation $\gamma_s$ is compared to the mutual coupling $J$, because of the nonzero $c_1$ and $c_2$ values. 
The quantum discord at the threshold of two DOPOs is calculated with the approximation $f(x)\rightarrow 1+\log(x/2) (x\rightarrow \infty)$. 
In the limit $p\rightarrow 1$ and $j\rightarrow \infty$, 
we have the maximum quantum discord $\mathcal{D}\sim \frac{1}{2}\log\frac{3}{4}+f(\sqrt{\frac{3}{2}})\sim 0.22$. 

\subsection{Numerical simulation}

To test the validity of analytical results for a ferromagnetically coupled two-DOPO system, 
we performed a numerical simulation based on the positive-$P$, truncated Wigner, and truncated Husimi SDEs. 
The simulation was performed with a small two-photon absorption loss $B/\gamma_s=10^{-4}$ and coupling coefficient $j=7/3$. 
The fluctuations were calculated with the time average for a single trajectory, assuming ergodicity. 
The total simulation time was a period of $10^6/\gamma_s$. 
In the first time period of $t_f=10^4/\gamma_s$, the time average was not taken, 
and the pump excitation varies on time via $p(t)=p\sqrt{t/t_f}$. 
After the first period of $t_f$, the excitation was held constant to $p$, and the time average was taken.  
The time step was $\Delta t=10^{-3}/\gamma_s$, $\Delta t=5\times 10^{-5}/\gamma_s$, and $\Delta t=2.5\times 10^{-5}/\gamma_s$, 
respectively for positive-$P$, truncated Wigner and truncated Husimi SDEs. 

\begin{figure*}
\begin{center}
\includegraphics[width=15cm]{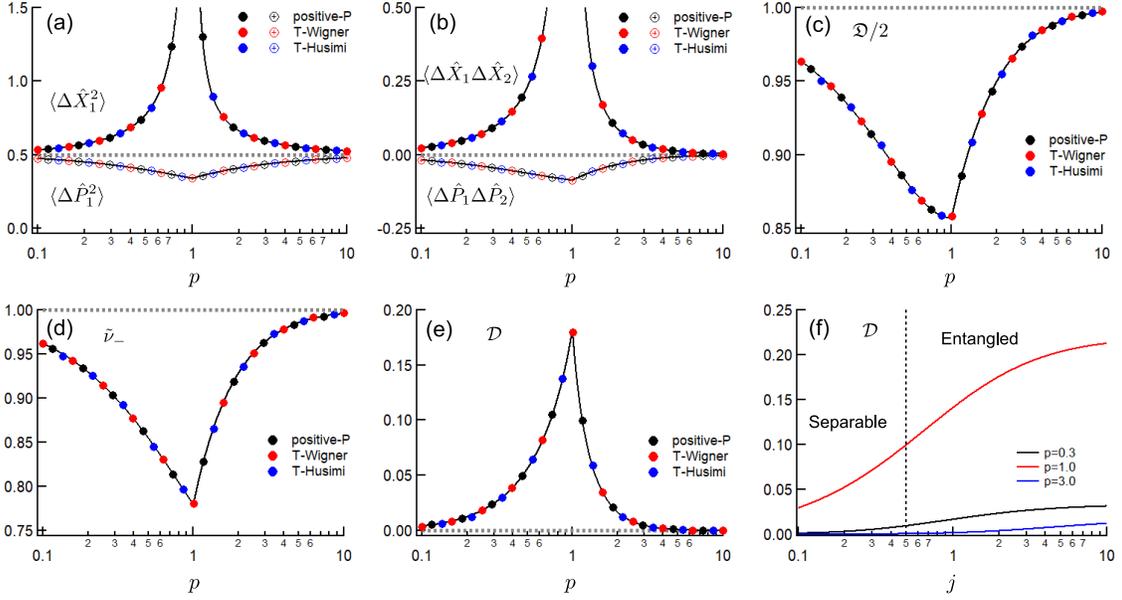}
\caption{Entanglement and quantum discord of two dissipatively coupled DOPOs. 
(a) Variances in canonical components $\langle \Delta \hat{X}^2\rangle$ (filled circles) and $\langle \Delta \hat{P}^2\rangle$ (open circles) as a function of the normalized excitation $p$. 
(b) Correlation of canonical components as a function of the normalized excitation $p$. 
(c) Duan-Giedke-Cirac-Zoller sufficient criterion for inseparability as a function of $p$. 
(d) Duan/Simon's necessary and sufficient criterion for inseparability as a function of $p$. 
(e) Quantum discord in coupled DOPOs as a function of $p$. 
(f) Analytical quantum discord as a function of normalized coupling coefficient $j=J/\gamma_s$. }
\label{2s}
\end{center}
\end{figure*}

We first present the results for $\langle \Delta \hat{X}^2\rangle=a_1/2$ 
and $\langle \Delta \hat{P}^2\rangle =a_2/2$ for the two-DOPO system in Fig.\ref{2s}(a). 
The circles with different colors represent the numerical results 
from positive-$P$, truncated-Wigner (T-Wigner), and truncated-Husimi (T-Husimi) calculations. 
The filled circles represent $\langle \Delta \hat{X}^2\rangle$, while the open circles represent $\langle \Delta \hat{P}^2\rangle$. 
The analytical results are shown by black solid lines. 
The fluctuation of vacuum is shown by a gray dashed line. 
Far below the threshold $(p\ll 1)$ and far above the threshold $(p\gg 1)$, 
the DOPO field should be in a vacuum state and a coherent state, respectively, so that the variance 
$\langle \Delta \hat{X}^2\rangle$ and $\langle \Delta \hat{P}^2\rangle$ are expected to approach an asymptotic value 
$\langle \Delta \hat{X}^2\rangle =\langle \Delta \hat{P}^2\rangle= \frac{1}{2}$. 
On the other hand, the degree of squeezing and anti-squeezing should be maximum at the threshold $(p=1)$. 
The numerical simulations reproduce these theoretical predictions well 
and thus verified the analytical results presented in the previous sections. 
The correlations of canonical coordinates are shown in Fig.\ref{2s}(b). 
Mean fluctuation product of canonical coordinates $\langle \Delta \hat{X}_1\Delta \hat{X}_2\rangle$ 
has a positive correlation and has a singular increase at the threshold, similarly to the $\langle \Delta \hat{X}^2\rangle$. 
The fluctuation product of canonical momenta $\langle \Delta \hat{P}_1\Delta \hat{P}_2\rangle$ has a negative correlation and remains finite at the threshold. 

Duan-Giedke-Cirac-Zoller sufficient criterion for entanglement ($\mathfrak{D}/2$) is shown in Fig.\ref{2s}(c). 
When this value is smaller than one, the system is inseparable and entangled. 
We see that the entanglement is maximum at the threshold, but exists even above the threshold. 
Note that the entanglement condition above the threshold ($j>1$) is satisfied in this numerical simulation. 
The theoretical values shown by black line were calculated from Eqs.(\ref{duanb}) and (\ref{duana}), 
and these analytical results match the numerical simulation results. 
The Duan-Simon's necessary and sufficient entanglement criterion is shown in Fig.\ref{2s}(d), 
which also shows a maximum entanglement at the threshold.  
Next, Fig.\ref{2s}(e) shows the quantum discord as a function of $p$, 
calculated from the covariance matrix $\sigma_{ij}$ \cite{Takata15}.  
The analytical results were obtained from Eqs.(\ref{qdc}) and (\ref{infs}) and are shown by a black line. 
As shown by filled circles, the quantum discord takes a maximum value at the threshold and remains finite even at the pump rate far below and far above it.  
Figure \ref{2s}(f) shows the analytical quantum discord of dissipatively coupled DOPOs 
at three different pump rates as a function of normalized dissipative coupling $j=J/\gamma_s$. 
A dashed vertical line represents the necessary and sufficient entanglement criterion for above threshold ($j=1/2$). 
For a large loss limit ($j\ll 1$), the entanglement disappears, but the quantum discord survives particularly around the threshold: $p\sim 1$. 

\subsection{Mean field approximation}

For comparison, we present here a theoretical model for mean-field coupled CIMs. 
A mean-field coupling theory in nonlinear quantum optical system has been developed 
in Bose-Hubbard like models \cite{Boite13,Savona17}. 
In such an approach, the Hamiltonian coupling between different sites, $\hat{H}_C=\hbar t\sum_{\langle ij\rangle}(\hat{a}_i^{\dagger}\hat{a}_j+\hat{a}_j^{\dagger}\hat{a}_i)$, 
is approximated by the mean-field via $\hat{a}_i^{\dagger}\hat{a}_j \rightarrow \hat{a}_i^{\dagger}\langle \alpha_j\rangle +\langle \alpha_i\rangle^*\hat{a}_j$. 
Here, $t$ is a hopping element. 
Similarly, we can consider the mean-field coupling in two dissipatively coupled DOPOs, 
where the Liouvillian coupling [Eq.(\ref{lc})] is replaced by 
$[\hat{a}_{1}-\hat{a}_{2},\hat{\rho}(\hat{a}_{1}^{\dagger}-\hat{a}_{2}^{\dagger})]\rightarrow [\hat{a}_{1}-\langle \alpha_{2}\rangle ,\hat{\rho}(\hat{a}_{1}^{\dagger}-\langle \alpha_{2}\rangle^{*})]+[\hat{a}_{2}-\langle \alpha_{1}\rangle ,\hat{\rho}(\hat{a}_{2}^{\dagger}-\langle \alpha_{1}\rangle^{*})]$. 
Assuming that $\langle \alpha_r \rangle (r=1,2)$ are real, $\mathcal{L}_C\hat{\rho}$ in Eq.(\ref{lc}) is replaced as 
\begin{eqnarray}
\label{mfa}
\mathcal{L}_C\hat{\rho}&=&J\sum_{r=1,2}([\hat{a}_r,\hat{\rho}\hat{a}_r^{\dagger}]+{\rm h.c.})\nonumber \\
&+&J\langle \alpha_2\rangle[\hat{a}_1^{\dagger}-\hat{a}_1,\hat{\rho}]+J\langle \alpha_1\rangle[\hat{a}_2^{\dagger}-\hat{a}_2,\hat{\rho}]. 
\end{eqnarray}
Such a model leads to a covariance matrix Eq.(\ref{cov2}) replaced by 
\begin{equation}
\label{covmf}
\underline{\alpha}=\Bigl[ 1+\frac{p}{1-p+j}, 1-\frac{p}{1+p+j} \Bigr], 
\end{equation}
below the threshold and by 
\begin{equation}
\label{covmf2}
\underline{\alpha}=\Bigl[ 1+\frac{1}{2(p-1)+j}, 1-\frac{1}{2p+j} \Bigr], 
\end{equation}
above the threshold. $c_1=c_2=0$ is satisfied both below and above the threshold in the mean field approximation. 
Far below the threshold ($p \ll 1$) and far above the threshold ($p\gg 1$), 
these covariance matrices are identical to those without mean field approximation [Eqs.(\ref{alphab})(\ref{gammab})(\ref{alphaa}) and (\ref{gammaa})], 
in the small coupling limit ($j \ll 1$). 
However, the difference emerges near the threshold ($p=1$).  
In the mean field approximation, $\langle \Delta \hat{X}_1^2\rangle$ is finite and $\langle \Delta \hat{X}_1\Delta \hat{X}_2\rangle$ is zero at the threshold, 
although they diverge without mean field approximation. 
The mean field coupled model has zero quantum discord, for $\inf_{\underline{\sigma_0}} \det \underline{\varepsilon}=\nu_+^2=\nu_-^2=a_1a_2$. 

\begin{figure*}
\begin{center}
\includegraphics[width=15cm]{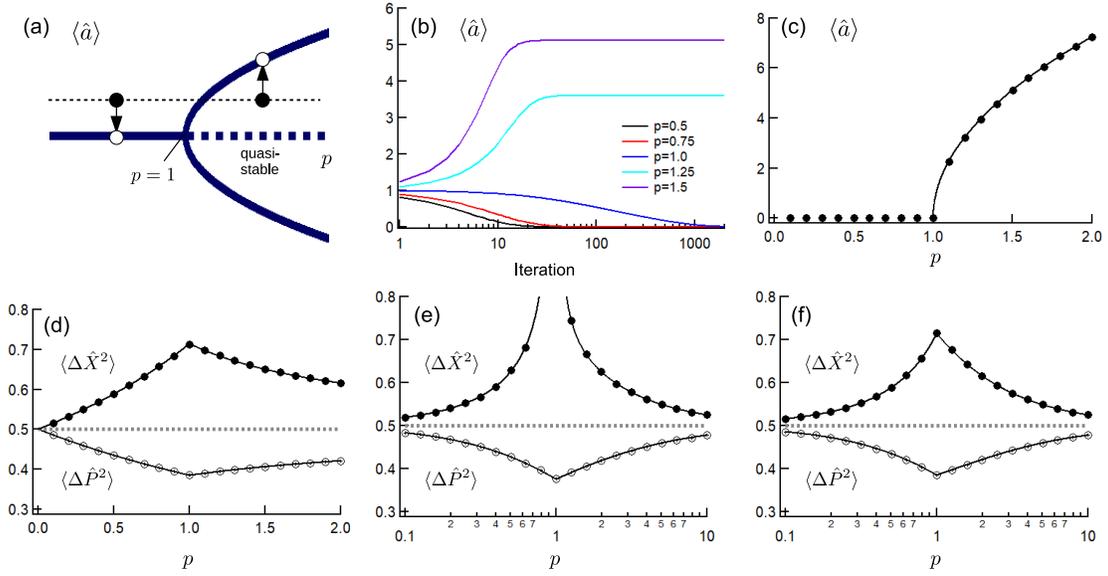}
\caption{Characteristics of mean-field coupled two DOPOs. 
(a) Diagram of bifurcation. 
(b) Calculation of self-consistent loop. 
(c) Mean amplitude by self-consistent loop as a function of $p$. 
(d) Fluctuations by self-consistent loop as a function of $p$. 
(e) Fluctuations by positive-$P$ calculation with $N_p=5$ particles. 
Black lines are obtained from Eqs.(\ref{dx2Np})(\ref{dp2Np}). 
(f) Fluctuations by positive-$P$ calculation with $N_p=9\times 10^4$ particles. 
Black lines are obtained from Eqs.(\ref{covmf})(\ref{covmf2}). }
\label{mfafig}
\end{center}
\end{figure*}

Mean field characteristics is calculated by the analytical results of the single DOPO\cite{Drummond81,Milburn81} 
and self-consistent loop\cite{Boite13}. 
The mean amplitude is 
\begin{equation}
\label{scl}
\langle \hat{a}\rangle=\frac{\sum_{k=0}^{\infty}\frac{2^k c^{2k+1} {}_2F_1(-k,x+e;2x;2) _2F_1(-k-1,x+e;2x;2)}{k!} }{\sum_{k=0}^{\infty}\frac{2^k c^{2k}{}_2F_1(-k,x+e;2x;2) _2F_1(-k,x+e;2x;2)}{k!} },
\end{equation}
where $c=\sqrt{\frac{S}{B}}$, $x=\frac{\gamma_s+J}{B}$, and $e=\frac{J\langle \hat{a}\rangle}{\sqrt{SB}}$.
The generalized moment is obtained by 
\begin{equation}
\langle \hat{a}^{\dagger m}\hat{a}^n\rangle=\frac{\sum_{k=0}^{\infty}\frac{2^k c^{k'+k''} {}_2F_1(-k',x+e;2x;2) _2F_1(-k'',x+e;2x;2)}{k!} }{\sum_{k=0}^{\infty}\frac{2^k c^{2k}{}_2F_1(-k,x+e;2x;2) _2F_1(-k,x+e;2x;2)}{k!} }, 
\end{equation}
where $k'=k+m$ and $k''=k+n$. 
We present the characteristics of mean-field coupled coherent Ising machines via self-consistent equation. 
In Fig.\ref{mfafig}(a), the bifurcation diagram is presented. 
When we consider the initial mean amplitude $\langle \hat{a}\rangle > 0$ (shown with filled circles in Fig.\ref{mfafig}(a)), 
from the self-consistent loop [Eq.(\ref{scl})], 
the mean amplitude is expected to converge onto $\langle \hat{a}\rangle=0$ below the threshold, 
and $\langle \hat{a}\rangle\sim \sqrt{\frac{\gamma_s}{B}(p-1)}$ above the threshold, shown with open circles in Fig.\ref{mfafig}(a). 
With the self-consistent loop, non-gaussian correction is precisely considered. 
We considered the results with $B/\gamma_s=0.019$ and $j=7/3$, starting from $\langle \hat{a}\rangle=1$. 
In Fig.\ref{mfafig}(b), we present the mean amplitude depending on the count of iteration in the self-consistent loop. 
For $p\le 1$, the mean amplitude converges into zero, while for $p>1$ the mean amplitude converges into finite values. 
In Fig.\ref{mfafig}(c), the mean amplitudes after $2000$ iterations were presented as a function of $p$. 
The mean amplitudes above the threshold are described by $\sqrt{\frac{\gamma_s}{B}(p-1)}$ shown by black line. 
The fluctuations are shown as circles in Fig.\ref{mfafig}(d). 
These fluctuations follow the black lines given by Eqs.(\ref{covmf})(\ref{covmf2}). 

Finally, we present the fluctuations by finite-particle simulation using positive-$P$ SDE. 
When mean field is constructed from $N_p$ positive-$P$ amplitudes, 
below the threshold steady-state fluctuations are 
\begin{eqnarray}
\label{dx2Np}
\langle \Delta \hat{X}_1^2\rangle &=& \frac{1}{2}+\frac{p}{2(1-p+j)} \\
&+&\frac{pj^2}{2N_p(1-p)(1-p+2j)(1-p+j)},\nonumber \\
\label{dx2xNp}
\langle \Delta \hat{X}_1 \Delta \hat{X}_2\rangle &=& \frac{pj}{2N_p(1-p)(1-p+2j)}, \\
\label{dp2Np}
\langle \Delta \hat{P}_1^2\rangle &=& \frac{1}{2}-\frac{p}{2(1+p+j)} \\
&-&\frac{pj^2}{2N_p(1+p)(1+p+2j)(1+p+j)}, \nonumber \\
\label{dp2xNp}
\langle \Delta \hat{P}_1 \Delta \hat{P}_2\rangle &=& -\frac{pj}{2N_p(1+p)(1+p+2j)}.
\end{eqnarray}
Above the threshold, steady state fluctuations of $\hat{X}$ are represented by Eqs.(\ref{dx2Np})(\ref{dx2xNp})
with substitution $p\rightarrow 1$ in the numerator and $1-p\rightarrow 2(p-1)$ in the denominator, 
and those of $\hat{P}$ are represented by Eqs.(\ref{dp2Np})(\ref{dp2xNp}) 
with substitution $p\rightarrow 1$ in the numerator and $1+p\rightarrow 2p$ in the denominator. 
These converge into the values of gaussian theory [Eqs.(\ref{covmf})(\ref{covmf2})] in the $N_p\rightarrow \infty$ limit. 
Fig.\ref{mfafig}(e)(f) show the positive-$P$ numerical results for $B/\gamma_s=10^{-4}$ and $j=7/3$,  
using $N_p=5$ and $9\times 10^4$ particles, respectively.  
Numerical simulation was performed in the similar way as Fig.\ref{2s}, 
where we considered $t_f=10^3/\gamma_s$ and total time development with period $2\times 10^3/\gamma_s$. 
For $N_p=5$, numerical results are well fitted by Eqs.(\ref{dx2Np})(\ref{dp2Np}), considering the correction of finiteness of particle number $N_p$. 
For $N_p=9\times 10^4$, numerical results are fitted well by Eqs.(\ref{covmf})(\ref{covmf2}). 
At $p=1$, the divergence of $\langle \Delta \hat{X}^2\rangle$ 
remaining in Eqs.(\ref{dx2Np})(\ref{dp2Np}) cannot be observed in Fig.\ref{mfafig}(f) due to large $N_p$ and finite $\gamma_s/B=10^4$.  

\section{1D lattice of $N$ DOPOs}

\subsection{Model}

We consider a CIM consisting of $N$-DOPOs ($\hat{a}_{r}(r=1,\cdots,N)$) which is the ferromagnetic 1D lattice with nearest neighbor coupling.  
The dissipative coupling Liouvillian is obtained from the extension of Eq.(\ref{lc}) as follows: 
\begin{equation}
\label{lc1dnn}
\mathcal{L}_C\hat{\rho}=\sum_{r=1}^{N} \frac{J}{2}[\hat{a}_{r}-\hat{a}_{r+1},\hat{\rho}(\hat{a}_{r}^{\dagger}-\hat{a}_{r+1}^{\dagger})]+{\rm h.c.}. 
\end{equation}
Here, we consider a periodic system such that $\hat{a}_{N+1}$ is identical to $\hat{a}_1$.  
Such a 1D lattice can be constructed with $N$ dissipative standing-wave modes $\hat{a}_{C,r}(r=1,\cdots,N)$ with half-width $\gamma_C$ : 
\begin{equation}
\hat{H}_C= \hbar \chi \sum_{r=1}^N (\hat{a}_r^{\dagger}-\hat{a}_{r+1}^{\dagger})\hat{a}_{C,r}+{\rm h.c.}. 
\end{equation}
Here, the $r$-th DOPO is coupled to the $(r-1)$-th DOPO via the dissipative mode $\hat{a}_{C,r-1}$, 
and is also coupled to the $(r+1)$-th DOPO via the dissipative mode $\hat{a}_{C,r}$.
When all dissipative modes $\hat{a}_{C,r}$ have the same loss rate $\gamma_C$ with $\gamma_C \gg \gamma_s$, 
the coupling coefficient of the Liouvillian follows $J/2 \sim \chi^2/\gamma_C$. 
In the Wigner and Husimi SDE, from Eq.(\ref{lc1dnn}), the coupling part in the SDE for the $r$-th DOPO is represented as follows: 
\begin{equation}
\label{wig1dnn}
\frac{d\alpha_r}{dt}|_C=-J\alpha_r+\frac{J}{2}(\alpha_{r-1}+\alpha_{r+1})+\sqrt{\frac{AJ}{2}}\xi_{C,r}-\sqrt{\frac{AJ}{2}}\xi_{C,r+1}. 
\end{equation}
Here, $A=1$ for the Husimi SDE and $A=\frac{1}{2}$ for the Wigner SDE. 
The noise sources satisfy $\langle \xi_{C,r}^*(t)\xi_{C,r'}(t')\rangle=2\delta_{r,r'} \delta (t-t')$. 
This model was obtained as the extension of the standing wave model in Fig.\ref{model1}. 
The theory of the traveling pulse model of a CIM \cite{Marandi14} in the Wigner representation is shown in Appendix C. 

\subsection{Extended Duan-Giedke-Cirac-Zoller criterion}

First, we take the analytical approach using the positive-$P$ SDE. 
For the $r$-th signal mode $\hat{a}_r(r=1,\cdots,N)$, the positive-$P$ SDE is written as:
\begin{eqnarray}
\frac{d\alpha_r}{dt}=-(\gamma_s+J) \alpha_r +\frac{J}{2}(\alpha_{r-1}+\alpha_{r+1}) \nonumber \\
+S\alpha_r^{\dagger}-B\alpha_r^{\dagger}\alpha_r^2 +\sqrt{S-B\alpha_r^2}\xi_{R,r}. 
\end{eqnarray}
We introduce the Fourier-transformed fluctuation amplitude $\Delta \tilde{\alpha}_k$: 
\begin{equation}
\Delta \alpha_{1+r}=\frac{1}{\sqrt{N}}\sum_{k=0}^{N-1}\Delta \tilde{\alpha}_k e^{i\theta_k r}. 
\end{equation}
Here, $\theta_k=\frac{2\pi}{N}k$.
Below the threshold, we obtain the following SDE for the Fourier components:
\begin{equation}
\frac{d\Delta \tilde{\alpha}_k}{dt}=-(\gamma_s+J(1-\cos\theta_k))\Delta \tilde{\alpha}_k+S\Delta \tilde{\alpha}_{-k}^{\dagger}+\sqrt{S}\tilde{\xi}_{R,k}. 
\end{equation}
The Fourier-transformed random noise sources are defined as 
\begin{equation}
\tilde{\xi}_{R,k}=\frac{1}{\sqrt{N}}\sum_{r=0}^{N-1} \xi_{R,r+1}e^{-i\theta_k r}, 
\end{equation} 
and satisfy $\langle \tilde{\xi}_{R,k}(t)\tilde{\xi}_{R,-k'}(t')\rangle=\delta_{k,k'}\delta(t-t')$. 
We obtain the steady-state fluctuation of Fourier components as follows.  
\begin{eqnarray}
\label{foumb}
\langle \Delta \tilde{\alpha}_k \Delta \tilde{\alpha}_{-k} \rangle &=& \frac{p}{2}\frac{1+j(1-\cos \theta_k)}{(1+j(1-\cos \theta_k))^2-p^2}, \\
\label{founb}
\langle \Delta \tilde{\alpha}_k^{\dagger}\Delta \tilde{\alpha}_k\rangle &=& \frac{p^2}{2}\frac{1}{(1+j(1-\cos \theta_k))^2-p^2}. 
\end{eqnarray}
We now calculate the extended Duan-Giedke-Cirac-Zoller criterion for even number $N$-DOPOs \cite{Maruo16}, 
denoted as $\mathfrak{D}'=\langle \Delta \hat{u}^2\rangle +\langle \Delta \hat{v}^2\rangle$, where 
$\hat{u}=\sum_{r=1}^N (-1)^{r+1} \hat{X}_r$ and $\hat{v}=\sum_{r=1}^N \hat{P}_r$. 
When $\mathfrak{D}'/N<1$, the entire system is not separated into the product state, i.e. $\hat{\rho} \ne \otimes_{r=1}^N \hat{\rho}_r$. 
The extended criterion can be written with Fourier components as 
\begin{eqnarray}
\label{Duanx}
\frac{\mathfrak{D}'}{N}=1+\langle \Delta \tilde{\alpha}_{k=0}^{\dagger}\Delta \tilde{\alpha}_{k=0}\rangle-\langle \Delta \tilde{\alpha}_{k=0}^2\rangle \nonumber \\
+\langle \Delta \tilde{\alpha}_{k=\frac{N}{2}}^{\dagger}\Delta \tilde{\alpha}_{k=\frac{N}{2}}\rangle+\langle \Delta \tilde{\alpha}_{k=\frac{N}{2}}^2\rangle. 
\end{eqnarray}
We can see that this expression is reduced to $\frac{\mathfrak{D}'}{N}=1-\frac{p(j-p)}{(1+p)(1-p+2j)}$. 
Note that this $\frac{\mathfrak{D}'}{N}$ contains the case of $N=2$ as a special case (see Eq.(\ref{duanb})). 

Above the threshold, we obtain the following SDE for the Fourier components:
\begin{eqnarray}
\label{daka}
\frac{d\Delta \tilde{\alpha}_k}{dt}=-(\gamma_s+J(1-\cos\theta_k))\Delta \tilde{\alpha}_k+S\Delta \tilde{\alpha}_{-k}^{\dagger} \nonumber \\
-2B\langle \alpha\rangle^2 \Delta \tilde{\alpha}_k-B\langle \alpha\rangle^2 \Delta \tilde{\alpha}_{-k}^{\dagger}+\sqrt{S-B\langle \alpha\rangle^2} \tilde{\xi}_{R,k}. 
\end{eqnarray}
Here, $S-B\langle \alpha\rangle^2 \sim \gamma_s$ is satisfied above the threshold. 
A similar equation was obtained in the dissipative Bose Hubbard model with the nonlinear Kerr effect\cite{Boite14}. 
From Eq.(\ref{daka}), we obtain the steady-state fluctuation correlations as follows. 
\begin{eqnarray}
\label{fouma}
\langle \Delta \tilde{\alpha}_k\Delta \tilde{\alpha}_{-k}\rangle &=& \frac{1}{2}\frac{2p-1+j(1-\cos \theta_k)}{(2p-1+j(1-\cos \theta_k))^2-1}, \\
\label{founa}
\langle \Delta \tilde{\alpha}_k^{\dagger}\Delta \tilde{\alpha}_k\rangle &=& \frac{1}{2}\frac{1}{(2p-1+j(1-\cos \theta_k))^2-1}. 
\end{eqnarray}
The extended Duan-Giedke-Cirac-Zoller criterion $\mathfrak{D}'/N$ is given by Eq.(\ref{Duanx}), 
and the result is reduced to $\frac{\mathfrak{D}'}{N}=1-\frac{j-1}{4p(p-1+j)}$, 
which is identical to $\mathfrak{D}/2$ [Eq.(\ref{duana})] for two DOPOs. 
Therefore, $j=1$ is required to satisfy the extended entanglement criterion above the threshold.

\begin{figure}
\begin{center}
\includegraphics[width=9cm]{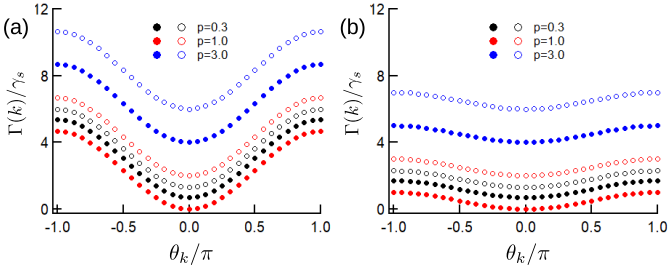}
\caption{Wave-number-dependent loss of $\hat{X}$-like mode (filled circles) and $\hat{P}$-like mode (open circles) 
in $N=32$ 1D ring of DOPOs with (a) $j=7/3$ and (b) $j=2/3$. 
When the loss of $\hat{P}$-like mode at $k=0$ is smaller than $\hat{X}$-like mode at $k=N/2$, 
the extended entanglement criterion is satisfied ($\mathfrak{D}'<N$).  
}
\label{lsp}
\end{center}
\end{figure}

To understand these results, we introduce an intuitive picture based on the loss spectrum model. 
After linearizaton, we consider the loss matrix $\underline{M_k}$ for fluctuation defined as 
\begin{equation}
\frac{d\langle \overrightarrow{A_k} \rangle}{d(\gamma_s t)}=-\underline{M_k}\langle \overrightarrow{A_k}\rangle, 
\end{equation}
where $\overrightarrow{A_k}:=\begin{bmatrix}\Delta \tilde{\alpha}_k \\ \Delta \tilde{\alpha}_{-k}^{\dagger} \end{bmatrix}$. 
The matrix $\underline{M_k}$ in a 1D DOPO system with nearest-neighbor coupling is described by
\begin{equation} 
\underline{M_k}=\begin{bmatrix} 1+j(1-\cos\theta_k)  & -p \\ -p & 1+j(1-\cos \theta_k)\end{bmatrix}, 
\end{equation}
below the threshold, and by 
\begin{equation}
\underline{M_k}=\begin{bmatrix} 2p-1+j(1-\cos\theta_k) & -1 \\ -1 & 2p-1+j(1-\cos\theta_k) \end{bmatrix} 
\end{equation}
above the threshold. 
When the loss matrix is diagonalized, 
the $\hat{X}$-like ($\Delta \tilde{\alpha}_k+\Delta \tilde{\alpha}_{-k}^{\dagger}$) mode has a normalized loss represented by 
$\Gamma_X(k)/\gamma_s=1-p+j(1-\cos\theta_k)$ below threshold and 
$\Gamma_X(k)/\gamma_s=2(p-1)+j(1-\cos\theta_k)$ above threshold. 
Likewise, the $\hat{P}$-like ($\Delta \tilde{\alpha_k}-\Delta \tilde{\alpha}_{-k}^{\dagger}$) mode 
has a normalized loss spectrum of $\Gamma_P(k)/\gamma_s=1+p+j(1-\cos\theta_k)$ below threshold and 
$\Gamma_P(k)/\gamma_s=2p+j(1-\cos\theta_k)$ above threshold. 
Figure \ref{lsp}(a) shows these spectra with $j=7/3$ for three excitation strengths $p=0.3$, $p=1$, and $p=3$. 
Here, the filled circles represent $\hat{X}$-like modes, while the open circles represent $\hat{P}$-like modes. 
As the fluctuation term does not depend on $\theta_k$ and the mode type (i.e., $\hat{X}$-like or $\hat{P}$-like ), 
the fluctuation magnitude is represented by the loss function. 
When the loss is smaller, the fluctuation is larger in proportion to $\Gamma(k)^{-1}$. 
We can see that when $p=1$, the $\hat{X}$-like mode becomes lossless at $k=0$ and lasing occurs at $k=0$. 
The extended entanglement criterion for the entire lattice is represented by fluctuations at two 
symmetric points in wavenumber space [Eq.(\ref{Duanx})]. 
The extended entanglement criterion ($\mathfrak{D}'/N$) is satisfied when the 
loss of the $\hat{X}$ -like band at $k=\frac{N}{2}$ is 
larger than that of the $\hat{P}$-like band at $k=0$. 
This is always the case far below the threshold. 
Above the threshold, however, the criterion is satisfied only when $j>1$. 
In Fig.\ref{lsp}(b), we present the wavenumber dependent loss for $j=2/3$. 
In this case, the extended entanglement criterion is satisfied for $p=0.3$, but not satisfied for $p=1$ and $p=3$. 

\begin{figure*}
\begin{center}
\includegraphics[width=15cm]{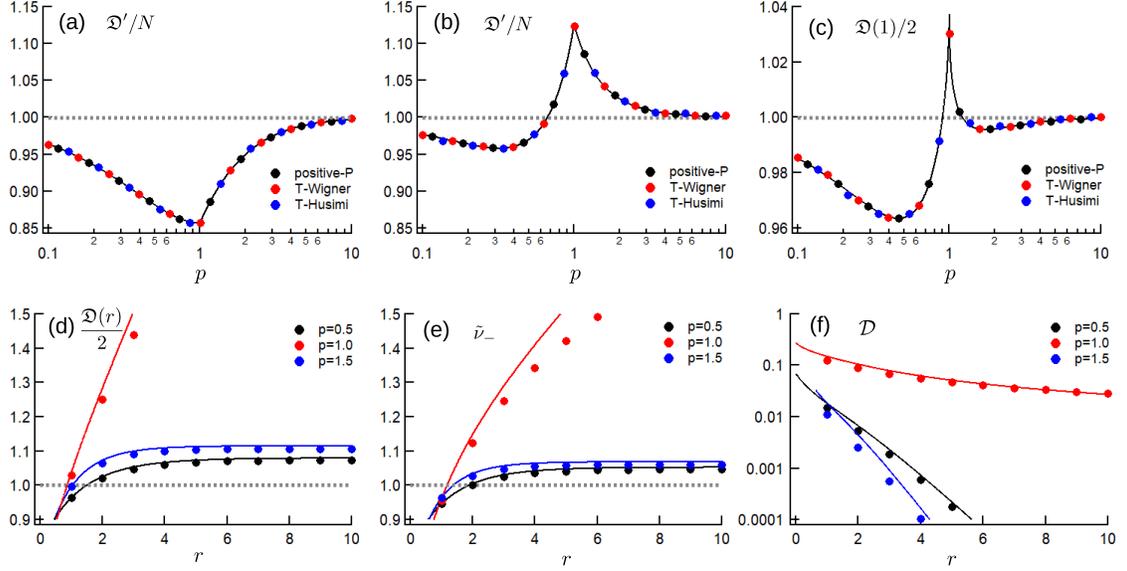}
\caption{Entanglement and quantum discord characteristics of $N$-DOPOs 
with 1D nearest-neighbor ferromagnetic coupling. 
Numerical result of extended entanglement criterion for $N=32$ lattice as a function of $p$ with (a) $j=7/3$ or (b) $j=2/3$. 
(c) Numerical result of Duan-Giedke-Cirac-Zoller sufficient inseparability criterion for a nearest-neighbor pair as a function of $p$ ($N=32$, $j=7/3$). 
Gray dashed lines represent the value of vacuum state, below which the nearest-neighbor DOPOs are inseparable. 
In (a-c), black lines represent the theoretical result with only gaussian approximation. 
(d) Duan-Giedke-Cirac-Zoller sufficient inseparability criterion as a function of site distance $r$ ($j=7/3$). 
(e) Duan/Simon necessary and sufficient inseparability criterion as a function of site distance $r$ ($j=7/3$). 
(f) Quantum discord as a function of site distance $r$ ($j=7/3$). 
In (d)(e) and (f), filled circles represent results from positive-$P$ calculation and lines represent results from parabolic approximation with $N\rightarrow \infty$. }
\label{1d}
\end{center}
\end{figure*}

We performed a numerical simulation to confirm the analytical results for the 1D DOPO ring with nearest-neighbor coupling. 
The simulation was performed in the same way as Fig.\ref{2s}(a-e), 
with a small two-photon absorption loss $B/\gamma_s=10^{-4}$. 
We considered $N=32$ DOPOs with $j=7/3$ or $j=2/3$. 
Figs.\ref{1d}(a)(b) show the extended entanglement criterion for an entire lattice. 
As seen in the loss spectrum of Fig.\ref{lsp}, the extended entanglement criterion is always satisfied for $j=7/3$. 
However, for $j=2/3$ it is satisfied only for $p<2/3$. 

\subsection{Entanglement criterion between nearest-neighbor DOPOs}

Instead of the entire lattice, we can consider the entanglement for two DOPOs, after tracing out the remaining $(N-2)$ DOPOs. 
Duan-Giedke-Cirac-Zoller sufficient entanglement criterion for a pair with site distance $r$ is calculated as follows:
$\frac{\mathfrak{D}(r)}{2}=1+\frac{2}{N}\sum_k(\langle \Delta \tilde{\alpha}^{\dagger}_k \Delta \tilde{\alpha}_k\rangle-\langle \Delta \tilde{\alpha}_{-k} \Delta \tilde{\alpha}_k \rangle \cos (r \theta_k))$. 
We first consider the $r=1$ case (nearest neighbor pair). 
In the limit $N\rightarrow \infty$, we compute the above summation by replacing it with the 
integration $\frac{1}{N}\sum_{k=0}^{N-1}f(\theta_k)\rightarrow \int_{-\pi}^{+\pi} f(\theta)\frac{d\theta}{2\pi}$. 
After the integration, we obtain 
\begin{equation}
\label{duanxb}
\frac{\mathfrak{D}(1)}{2}=1-\frac{p}{j}\Bigl[-1+\frac{1}{2}\sqrt{\frac{1-p}{1-p+2j}}+\frac{1}{2}\sqrt{\frac{1+p+2j}{1+p}}\Bigr] 
\end{equation}
below the threshold, and 
\begin{equation}
\label{duanxa}
\frac{\mathfrak{D}(1)}{2}=1-\frac{1}{j}\Bigl[-1+\frac{1}{2}\sqrt{\frac{p-1}{j+p-1}}+\frac{1}{2}\sqrt{\frac{j+p}{p}}\Bigr] 
\end{equation}
above it. We can see that $\mathfrak{D}(1)/2<1$ is always satisfied far below the threshold ($p\ll 1$), 
when $j>3$ satisfied at the threshold ($p\rightarrow 1$), and when $j>2$ satisfied far above the threshold ($p\gg 1$). 
When $2<j<3$, Duan-Giedke-Cirac-Zoller sufficient entanglement criterion 
is not satisfied at $p\sim 1$, even though it is satisfied well below and well above the threshold. 
Figure \ref{1d}(c) shows the Duan-Giedke-Cirac-Zoller sufficient entanglement criterion for a nearest-neighbor pair in $N=32$ DOPOs with $j=7/3$. 
The analytical results [Eqs.(\ref{duanxb}) and (\ref{duanxa})] agree very well with the numerical results. 
It is seen that the entanglement criterion ($\frac{\mathfrak{D}(r)}{2}<1$) is not satisfied at the threshold but is satisfied below and above it.  

\subsection{Correlation as a function of distance $r$}

Here, we consider the covariance matrix [Eq.(\ref{cov2})] between $\hat{a}_1$ and $\hat{a}_{1+r}$, 
calculated with $\overrightarrow{R}(r)=\sqrt{2}[\hat{X}_1,\hat{P}_1,\hat{X}_{1+r},\hat{P}_{1+r}]$. 
In the large $N$ limit, the small-$k$ part of the loss spectrum is important in calculating 
the spatial correlation function of the canonical components, 
$c_1=2\langle \Delta \hat{X}_1 \Delta \hat{X}_{1+r}\rangle$ and $c_2=2\langle \Delta \hat{P}_1 \Delta \hat{P}_{1+r}\rangle$. 
These are calculated from 
\begin{eqnarray}
c_1 &=& \frac{2}{N}\sum_{k} (\langle \Delta \tilde{\alpha}_k \Delta \tilde{\alpha}_{-k} \rangle+\langle \Delta \tilde{\alpha}_k^{\dagger}\Delta \tilde{\alpha}_k\rangle) \cos (\theta_k r), \\
c_2 &=& \frac{2}{N}\sum_{k} (-\langle \Delta \tilde{\alpha}_k \Delta \tilde{\alpha}_{-k} \rangle+\langle \Delta \tilde{\alpha}_k^{\dagger}\Delta \tilde{\alpha}_k\rangle) \cos (\theta_k r). 
\end{eqnarray}
In the limit $N\rightarrow \infty$, these expressions are reduced to the integration forms: 
\begin{eqnarray}
c_1 &= & \frac{\psi(p)}{\pi j} {\rm Re} \int_{-\pi}^{\pi} d\theta \frac{e^{i\theta r}}{\theta^2+m_-^2}, \\
c_2 &= & -\frac{\psi(p)}{\pi j} {\rm Re} \int_{-\pi}^{\pi} d\theta \frac{e^{i\theta r}}{\theta^2+m_+^2}, 
\end{eqnarray}
after the approximation $\cos \theta \sim 1-\frac{\theta^2}{2}$. 
Here, $\psi(p)=pH(1-p)+H(p-1)$ and $H(p)$ is the Heaviside's step function. 
$m_{\pm}^2=\frac{2(1\pm p)}{j}$ below the threshold and $m_{\pm}^2=\frac{2(2p-1\pm 1)}{j}$ above the threshold. 
Assuming small $m_{\pm}$, the covariance matrix can be calculated with complex integral as 
\begin{eqnarray}
\label{cov1d}
\underline{\alpha} &=& {\rm diag}\Bigl[1+\frac{\psi(p)}{jm_-},1-\frac{\psi(p)}{jm_+}\Bigr], \\
\label{cov1d2}
\underline{\gamma} &=& {\rm diag}\Bigl[\frac{\psi(p) e^{-m_- r}}{jm_-},-\frac{\psi(p) e^{-m_+ r}}{jm_+}\Bigr]. 
\end{eqnarray}

We see that the correlation length $r_c$ of the $\hat{X}$ component satisfies $r_c\propto |1-p|^{-\frac{1}{2}}$. 
The correlation of $\hat{X}$ becomes long range at the threshold. 
This result resembles the correlation length $r_c \propto |1-\frac{T}{T_c}|^{-\frac{1}{2}}$ in Landau's equilibrium phase transition theory \cite{Landau80}, 
where $T/T_c$ is a system temperature normalized by the critical temperature. 
The correspondence between an equilibrium phase transition and a non-equilibrium phase transition has been noted in Ref.\cite{DeGiorgio70}. 
For $\hat{P}$, on the other hand, the maximum correlation length is obtained at $p\rightarrow 0$: $r_c =\sqrt{j/2}$. 
Even for $p \rightarrow 0$, however, this correlation is short ranged. 
Duan-Giedke-Cirac-Zoller sufficient condition for entanglement for a pair with distance $r$ ($\mathfrak{D}(r)$) can be calculated 
from correlations shown above as 
\begin{equation}
\label{duanr}
\frac{\mathfrak{D}(r)}{2}=1+\frac{\psi(p)}{2j}\Bigl[\frac{1-e^{-m_- r}}{m_-}-\frac{1+e^{-m_+ r}}{m_+}\Bigr]. 
\end{equation}
Particularly, at the threshold ($p=1$), $\mathfrak{D}(r)$ is proportional to $r$ due to small $m_-$. 
With such dependence on $r$, the entanglement disappears for DOPO pair with large distance at the threshold. 

Considering both nearest neighbor value and the long-range characteristics, 
we obtain the equation, 
\begin{equation}
\label{nnc1}
a_1-c_1 = 1+\frac{\psi(p)}{j} \Bigl[ 1-\frac{m_-}{\sqrt{m_-^2+4}}-\frac{e^{-m_- r} -e^{-m_-}}{m_-} \Bigr], 
\end{equation}
\begin{equation}
\label{nnc2}
a_2+c_2 = 1+\frac{\psi(p)}{j} \Bigl[ 1-\frac{\sqrt{m_+^2+4}}{m_+}-\frac{e^{-m_+ r} -e^{-m_+}}{m_+} \Bigr]. 
\end{equation}
The theoretical Duan-Giedke-Cirac-Zoller sufficient condition for entanglement between DOPOs 
with site distance $r$ is shown in Fig.\ref{1d}(d), where $j=7/3$. 
Lines represent the values with complex integral and exact nearest-neighbor values[Eq.(\ref{nnc1})(\ref{nnc2})], and 
filled circles are calculated from positive-$P$ SDEs with $N=32$. 
Duan-Giedke-Cirac-Zoller entanglement criterion is satisfied only for $r=1$ even with $p\ne 1$. 
In Fig.\ref{1d}(e), necessary and sufficient condition for entanglement is presented. 
Entanglement criterion is satisfied even for nearest-neighbor pair for $p=1$. 
However, for pairs with large $r$, entanglement criterion is not satisfied. 
In Fig.\ref{1d}(d)(e), numerical results obtained by positive-$P$ calculation are shown with filled circles, 
which match the analytical results. 
At the threshold $p\rightarrow 1$, difference between the theory and numerical results emerges 
for large $r$ presumably because analytical results do not assume periodicity of one-dimensional lattice. 

We next consider the quantum discord between two DOPOs with site distance $r$. 
The theoretical quantum discord between DOPOs with site distance $r$ is shown with lines in Fig.\ref{1d}(f), where $j=7/3$. 
Lines represent the values calculated with Eqs.(\ref{cov1d})(\ref{cov1d2}).  
Filled circles represent the numerical results obtained by positive-$P$ simulation. 
Even for large distance $r$, non-zero quantum discord can be formed at the threshold. 
This is because $c_1$ has a long-range correlation at the threshold. 
Even when $c_2$ is negligibly small for large $r$, the non-zero quantum discord can be obtained via finite $c_1$ value. 

\section{Summary}

We have presented analytical results on the degrees of squeezing/anti-squeezing, entanglement and 
quantum discord for a network of DOPOs both below and above threshold. 
We confirmed the validity of the analytical results through numerical simulations 
based on the positive-$P$, truncated Wigner, and truncated Husimi SDEs. 
The comparison establishes well the validity of the phase-space methods based on these SDEs. 
The $c$-number Langevin equations are essentially identical to 
the truncated Wigner SDE and valid for DOPOs below and above threshold. 
In a high-loss CIM based on optical delay-line coupling, 
entanglement between DOPO pulses disappears before reaching the threshold\cite{Takata15}. 
It does, however, have quantum discord, indicating the existence of quantum correlations 
among the quantum fluctuations of DOPO pulses below and above threshold. 
Such quantum correlations disappear if we use a mean-field coupling 
that does not account for the fluctuations in the coupling fields. 
In the one-dimensional lattice of DOPOs, the entanglement criterion 
is satisfied only for short ranged pair of DOPOs. 
The quantum discord of far distant pair, on the other hand, 
has finite values particularly around the threshold, due to the long range correlation of canonical coordinates. 

The analytical methods and insights obtained here for a 1D lattice of DOPOs with nearest-neighbor coupling seem to be 
applicable to other systems: DOPOs with measurement feedback coupling\cite{Shoji17}, 
more complicated DOPO lattices with frustration\cite{Hamerly16,Leleu17} or topological nontriviality of the loss spectrum, 
and optical networks with other nonlinear components. 
For example, in the coherent XY machine\cite{Takeda17} with dissipative coupling, the entanglement has not been considered. 
Some lasers, for example Raman lasers, are known to have 
photon number squeezed state (sub-Poissonian state) only far above the threshold\cite{Yamamoto86,Ritsch92}. 
As the dissipatively coupled squeezed states can satisfy entanglement criterion\cite{Takata15,Maruo16}, 
the coherent XY machines could also satisfy entanglement criterion above the threshold, 
due to the dissipative coupling of the photon number squeezed states. 

The linearization treatment developed in this paper is effective only for weak noise limit. 
They will not be applied with sufficient accuracy for large quantum noise case, 
where non-gaussianity \cite{Yamamura17,Drummond89} becomes important. 

\begin{acknowledgments}
The authors thank Z.Wang, T.Leleu, K.Takata, P.D.Drummond, Y.Huang, H.Mabuchi and E.Ng for their critical discussions. 
This work was supported partially by the ImPACT program of the cabinet office of the government of Japan. 
\end{acknowledgments}

\appendix

\section{Phase-space method}

We expand the density operator in the quantum master equation (\ref{QME}) 
by using the positive-$P$, Wigner, and Husimi quasi-distribution functions. 
First, we discuss Glauber's diagonal $P$ distribution function, 
which is represented as $\hat{\rho}=\int P(\alpha)|\alpha\rangle \langle \alpha| d^2\alpha$\cite{Glauber63}. 
For a DOPO, the Fokker-Planck equation of the $P$ function is derived as : 
\begin{equation}
\label{gpfp}
\frac{\partial P}{\partial t}=\frac{\partial}{\partial \alpha}[(\gamma_s\alpha-S\alpha^*+B|\alpha|^2\alpha)P]+\frac{1}{2}\frac{\partial^2}{\partial \alpha^2}[(S-B\alpha^2)P]+{\rm c.c.} . 
\end{equation}
The diagonal $P$ function is not positive, however, and it does not have a corresponding $c$-number SDE 
because of the difficulty in adding noise term to $d\langle \alpha^2\rangle/dt$ without adding it to $d\langle |\alpha^2|\rangle /dt$.
The positive-$P$ representation\cite{Drummond80} is a modification to Glauber's $P$ representation and 
is known to have a similar Fokker-Planck equation.  
We can write the distribution function in positive-$P$ representation as \cite{Drummond80} 
\begin{equation}
\label{ppex}
\hat{\rho}=\int P(\alpha,\alpha^{\dagger})\frac{|\alpha\rangle \langle \alpha^{\dagger *}|}{\langle \alpha^{\dagger *}|\alpha\rangle}d^2\alpha d^2\alpha^{\dagger} . 
\end{equation} 
Here, $\alpha$ and $\alpha^{\dagger}$ are two independent complex numbers satisfying $\langle \alpha\rangle^*=\langle \alpha^{\dagger}\rangle$. 
Using Eq.(\ref{ppex}) for Eq. (\ref{QME}), we obtain the quantum-mechanical Fokker-Planck equation for $P(\alpha,\alpha^{\dagger})$. 
This Fokker-Planck equation for the positive-$P$ representation is the same as that for Glauber's diagonal $P$ distribution, 
except for the replacement $\alpha^*\rightarrow \alpha^{\dagger}$. 
Applying the Ito rule to the resulting Fokker-Planck equation, we can obtain the following $c$-number SDE for the signal mode\cite{Shoji17}: 
\begin{equation}
\label{pp1}
\frac{d\alpha}{dt}=-\gamma_s\alpha + S\alpha^{\dagger}-B\alpha^{\dagger}\alpha^2 + \sqrt{S-B\alpha^2}\xi_{R}, 
\end{equation}
\begin{equation}
\label{pp2}
\frac{d\alpha^{\dagger}}{dt}=-\gamma_s\alpha^{\dagger} + S\alpha-B\alpha^{\dagger 2}\alpha + \sqrt{S-B\alpha^{\dagger 2}}\xi_{R}^{\dagger}. 
\end{equation}
Here, $\xi_{R}$ and $\xi_{R}^{\dagger}$ are independent real-number random noise sources satisfying 
$\langle \xi_R(t)\xi_R(t')\rangle=\delta(t-t')$ and $\langle \xi_R^{\dagger}(t)\xi^{\dagger}_R(t') \rangle=\delta(t-t')$. 
This SDE under positive-$P$ representation is defined with no truncation and is equivalent to the quantum master equation (\ref{QME}). 
In numerical simulation, however, these equations have some difficulty for large $B/\gamma_s$. 
It is known that when $B$ is large, positive-$P$ calculation becomes unstable \cite{Gilchrist97,Deuar02}. 
This is because a negative photon number is allowed in positive-$P$ calculation, 
which causes the photon number to diverge to negative infinity because of two-photon absorption. 
In this paper, we thus consider the case of small $B$ and a sufficiently slow adiabatic excitation schedule. 
In our simulations, the divergence problem was not observed. 

As noted above, the SDEs with positive-$P$ representation are equivalent to the Liouville equation (\ref{QME}). 
On the other hand, 
the Wigner and the Husimi functions for Eq.(\ref{QME}) have third and higher order derivatives in their Fokker-Planck equations. 
As these higher order derivatives are proportional to $B/\gamma_s$, 
we can neglect higher-order derivatives when $B/\gamma_s$ is small, and we obtain truncated SDEs. 
In return for the possible incorrectness due to truncation, 
the truncated Wigner and Husimi SDEs represent the signal boson operator $\hat{a}$ by only one $c$-number amplitude. 
The Wigner distribution function is written with a density matrix as the following\cite{Cahill69,Mishkin69}: 
\begin{equation}
W(\alpha)=\frac{1}{\pi^2}\int {\rm tr}\hat{\rho} e^{\eta(\hat{a}^{\dagger}-\alpha^*)-\eta^*(\hat{a}-\alpha)}d^2\eta. 
\end{equation}
The Fokker-Planck equation can be obtained from the time derivative of this equation.  
The truncated Wigner SDE is then obtained from the Fokker-Planck equation after neglecting third-order derivatives and 
terms resulting from the Weyl ordering in the drift and diffusion terms with the assumption of $B\ll \gamma_s$ :  
\begin{equation}
\label{WSDE}
\frac{d\alpha}{dt}=-\gamma_s\alpha + S\alpha^{*}-B|\alpha|^2\alpha + \sqrt{\frac{\gamma_s}{2}+B|\alpha|^2}\xi_{C}. 
\end{equation}
We can see that this Wigner SDE resembles the $c$-number counterpart of the Heisenberg-Langevin equation after eliminating the pump mode \cite{Wang13}, 
but Eq.(\ref{WSDE}) is the $c$-number SDE in the Schr\"odinger picture rather than the $q$-number equation in the Heisenberg picture. 

The Husimi distribution function is then written as $Q(\alpha)=\frac{1}{\pi}\langle \alpha | \hat{\rho}|\alpha \rangle$. 
The Fokker-Planck equation of the Husimi distribution function for degenerate two-photon absorption is presented in \cite{Drummond81b}, 
and it has third- and fourth- order derivatives. 
Here, we consider the truncated Husimi Fokker-Planck equation of Eq.(\ref{QME}) only up to second-order derivatives. 
As this approximation assumes small $B/\gamma_s$, we also neglect small $B/\gamma_s$ dependent terms 
resulting from the anti-normal ordering of drift and diffusion terms. 
The truncated Husimi SDE is obtained through the Fokker-Planck equation as the following: 
\begin{eqnarray}
\label{HSDE}
\frac{d\alpha}{dt}&=&-\gamma_s\alpha + S\alpha^{*}-B|\alpha|^2\alpha+ \sqrt{\gamma_s-\frac{S}{2}+\frac{3}{2}B|\alpha|^2}\xi_{C}\nonumber \\
&+& i\sqrt{S}\xi_{R1}+\sqrt{B}\alpha\xi_{R2}. 
\end{eqnarray}
Here, $\xi_C$ is a complex-number noise source with $\langle \xi_C^{*}(t)\xi_C(t')\rangle=2\delta(t-t')$,  
and $\xi_{Ri}$ is a real-number noise source with $\langle \xi_{Ri}(t)\xi_{Rj}(t')\rangle=\delta_{ij}\delta (t-t')$. 
We can use Eqs.(\ref{pp1}) and (\ref{pp2}) for the first principle numerical simulation of a CIM, 
and Eqs.(\ref{WSDE}) and (\ref{HSDE}) for the approximated simulation of a CIM with $B\ll \gamma_s$. 

\section{Analytical method for single DOPO}

Here, we derive the analytical solutions for the quantum noise of a CIM. 
We use the positive-$P$ representation, 
while noting that the truncated Wigner and truncated Husimi representations can produce identical results in the weak noise limit ($B/\gamma_s\ll 1$). 
Below the threshold, we neglect the two-photon absorption terms with $B$ in Eqs.(\ref{pp1}) and (\ref{pp2}). 
We can then obtain the following positive-$P$ SDEs: 
\begin{equation}
\label{pp1b}
\frac{d\alpha}{dt}=-\gamma_s\alpha + S\alpha^{\dagger}+ \sqrt{S}\xi_{R}
\end{equation}
\begin{equation}
\label{pp2b}
\frac{d\alpha^{\dagger}}{dt}=-\gamma_s\alpha^{\dagger} + S\alpha+ \sqrt{S}\xi_{R}^{\dagger}
\end{equation}
The term with $B$ in the square root of Eqs.(\ref{pp1}) and (\ref{pp2}) can contribute even below the threshold when $B/ \gamma_s$ is large, 
as $d\langle \alpha^2\rangle/dt=-(2\gamma_s+B) \langle \alpha^2\rangle+\cdots$. 
When $B$ is large, the saturation term $-B\alpha^{\dagger}\alpha^2$ in Eq.(\ref{pp1}) is also not negligible even below the threshold, because 
the rate of two-photon absorption is enhanced by the intensity correlation function 
$g^{(2)}(0)=\frac{\langle \hat{a}^{\dagger 2}\hat{a}^2\rangle}{\langle \hat{a}^{\dagger}\hat{a}\rangle^2}$ \cite{Shen67}, 
and a squeezed vacuum state in a DOPO has large $g^{(2)}(0)$ proportional to $\frac{1}{2\langle \hat{a}^{\dagger}\hat{a}\rangle}$. 
We can neglect the saturation term, however, if $B\ll \gamma_s$. 
Below the threshold, the mean amplitude of the system is $\langle \alpha\rangle=0$. 
Using the normalized excitation $p=S/\gamma_s$, which is equal to $\varepsilon/\varepsilon_{thr}$, 
we can obtain the following steady-state equation: 
\begin{equation}
\label{1opobt}
\begin{bmatrix}
1 & -p \\
-p & 1
\end{bmatrix}
\begin{bmatrix}
\langle \Delta \alpha^2\rangle \\
\langle \Delta \alpha^{\dagger} \Delta \alpha \rangle
\end{bmatrix}
=\frac{p}{2}
\begin{bmatrix}
1\\ 0
\end{bmatrix}. 
\end{equation}
We assume that $\langle \Delta \alpha^2\rangle$ is real, because from Eqs.(\ref{pp1b}) and (\ref{pp2b}), $\alpha$ and $\alpha^{\dagger}$ are real: 
the time development starts from the vacuum state $\alpha=\alpha^{\dagger}=0$, and both the coefficients and random noises are real. 
We define the canonical coordinate and momentum as $\hat{X}=\frac{\hat{a}+\hat{a}^{\dagger}}{\sqrt{2}}$ 
and $\hat{P}=\frac{\hat{a}-\hat{a}^{\dagger}}{\sqrt{2} i}$, respectively. 
In the positive-$P$ representation, the fluctuations are calculated as 
$\langle \Delta \hat{X}^2\rangle=\langle \Delta \alpha^{\dagger}\Delta \alpha\rangle+\langle \Delta \alpha^2\rangle+\frac{1}{2}$ and 
$\langle \Delta \hat{P}^2\rangle=\langle \Delta \alpha^{\dagger}\Delta \alpha\rangle-\langle \Delta \alpha^2\rangle+\frac{1}{2}$. 
In a DOPO below the threshold under the steady-state condition, these canonical components have the following fluctuations: 
$\langle \Delta \hat{X}^2\rangle=\frac{1}{2}+\frac{p}{2(1-p)}$ and $\langle \Delta \hat{P}^2\rangle=\frac{1}{2}-\frac{p}{2(1+p)}$. 
We can see that $\langle \Delta \hat{X}^2\rangle$ has a singularity at the threshold. 
On the other hand, $\langle \Delta \hat{P}^2\rangle$ has a minimized value $\langle \Delta \hat{P}^2\rangle=\frac{1}{4}$, 
which is one half the vacuum fluctuation, at the threshold\cite{Milburn81}. 
Similar results can be obtained from other representations; 
for example, in the truncated Wigner representation, we set $B\rightarrow 0$ in Eq.(\ref{WSDE}). 
From 
$\begin{bmatrix}
1 & -p \\
-p & 1
\end{bmatrix}
\begin{bmatrix}
\langle \Delta \alpha^2\rangle \\
\langle |\Delta \alpha|^2 \rangle
\end{bmatrix}
=\frac{1}{2}
\begin{bmatrix}
0 \\ 1
\end{bmatrix}$, 
we obtain the same mean-squared fluctuations via $\langle \Delta \hat{X}^2\rangle=\langle |\Delta \alpha|^2 \rangle+\langle \Delta \alpha^2\rangle$ and 
$\langle \Delta \hat{P}^2\rangle=\langle |\Delta \alpha|^2 \rangle-\langle \Delta \alpha^2\rangle$. 

Next, we consider the fluctuations of a DOPO above the threshold. 
We separate the signal amplitudes into the mean amplitudes $\langle \alpha\rangle=\langle \alpha^{\dagger}\rangle= \sqrt{\frac{\gamma_s}{B}(p-1)}$ and 
small fluctuations ( $\Delta \alpha=\alpha-\langle \alpha\rangle $ and $\Delta \alpha^{\dagger}=\alpha^{\dagger}-\langle \alpha^{\dagger}\rangle$ ). 
Such linearization treatment was previously applied in quantum nonlinear optics 
for degenerate two-photon absorption \cite{Chaturvedi77} and an above-threshold DOPO\cite{McNeil78,Drummond79}. 
We obtain the positive-$P$ SDE for the fluctuation part as follows. 
\begin{eqnarray}
\frac{d \Delta \alpha}{dt}=-\gamma_s \Delta \alpha+S\Delta \alpha^{\dagger}-2B|\langle \alpha\rangle|^2 \Delta \alpha\nonumber \\
-B\langle \alpha\rangle^2 \Delta \alpha^{\dagger}+\sqrt{S-B\langle \alpha\rangle^2}\xi_{R},
\end{eqnarray}
\begin{eqnarray}
\frac{d \Delta \alpha^{\dagger}}{dt}=-\gamma_s \Delta \alpha^{\dagger}+S\Delta \alpha-2B|\langle \alpha\rangle|^2 \Delta \alpha^{\dagger}\nonumber \\
-B\langle \alpha^{\dagger}\rangle^2 \Delta \alpha+\sqrt{S-B\langle \alpha^{\dagger}\rangle^2}\xi_{R}^{\dagger}. 
\end{eqnarray}
Using the normalized excitation $p=\varepsilon/\varepsilon_{thr}$, we then obtain the following equation under the steady-state condition: 
\begin{equation}
\label{1opoat}
\begin{bmatrix}
2p-1 & -1 \\
-1 & 2p-1
\end{bmatrix}
\begin{bmatrix}
\langle \Delta \alpha^2\rangle \\
\langle \Delta \alpha^{\dagger} \Delta \alpha \rangle
\end{bmatrix}
=\frac{1}{2}
\begin{bmatrix}
1\\ 0
\end{bmatrix}. 
\end{equation}
The squared fluctuations of the canonical coordinate and momentum are obtained as 
$\langle \Delta \hat{X}^2\rangle=\frac{1}{2}+\frac{1}{4(p-1)}$ and $\langle \Delta \hat{P}^2\rangle=\frac{1}{2}-\frac{1}{4p}$, respectively. 
Above the threshold, the canonical coordinate and momentum monotonically reach the vacuum fluctuation. 
Although squeezing exists above the threshold, the squeezing becomes weaker for large $p$ \cite{Milburn81}. 
We can see that Eqs.(\ref{1opobt}) and (\ref{1opoat}) are reduced to an identical equation at the threshold (in the limit $p\rightarrow 1$). 
For the canonical momentum $\hat{P}$, the same mean-squared fluctuations are obtained 
at $p=1$ from both analytical theories below and above the threshold. 
For the canonical coordinate $\hat{X}$, the mean-squared fluctuation above the threshold 
is obtained from that below the threshold with substitution 
$p\rightarrow 1$ in the numerator and $1-p\rightarrow 2(p-1)$ in the denominator. 
For the canonical momentum $\hat{P}$, the mean-squared fluctuation above the threshold 
is obtained from that below the threshold with substitution 
$p\rightarrow 1$ in the numerator and $1+p\rightarrow 2p$ in the denominator. 

\section{Traveling-pulse model of 1D lattice}

\begin{figure}
\begin{center}
\includegraphics[width=8cm]{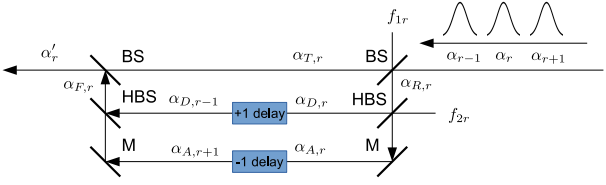}
\caption{Traveling pulse model of a 1D lattice of CIMs with nearest-neighbor coupling. }
\label{model2}
\end{center}
\end{figure}

We also consider the traveling-pulse model \cite{Marandi14} 
of the nearest-neighbor-coupled 1D lattice of DOPOs, shown in Fig.\ref{model2}. 
This model uses traveling optical pulses in a single ring cavity with a nonlinear crystal, 
in which the pulses are coupled with an optical delay-line. 
We assume that the two beam splitters (BSs) have reflectance $R_B$, 
the two half-beam splitters (HBSs) have $50\%$ reflectance, and 
the mirrors (Ms) have $100\%$ reflectance. 
The setting shown in Fig.\ref{model2} has HBSs in the delay-line and is, 
as we show below, equivalent to the Liouvillian coupling of Eq.(\ref{lc1dnn}). 
We note that some models \cite{Hamerly16,Inagaki16} discussed in earlier papers are not equivalent to Eq.(\ref{lc1dnn}) 
because they have no HBSs in the delay-line.  
The traveling-pulse model is easily implemented with the Heisenberg Langevin model or SDE for the Wigner representation. 

In the Wigner representation, the $r$-th signal pulse $\alpha_r$ is mixed with the vacuum noise $f_{1r}$ at a beam splitter (BS) 
and transformed into the transmitted $\alpha_{T,r}$ and reflected $\alpha_{R,r}$ modes: 
$\alpha_{T,r}=\sqrt{1-R_B}\alpha_r+\sqrt{R_B}f_{1r}$ and $\alpha_{R,r}=\sqrt{R_B}\alpha_r-\sqrt{1-R_B}f_{1r}$. 
The reflected light is then separated by an HBS into 
$\alpha_{D,r}=\frac{1}{\sqrt{2}}f_{2r}+\frac{1}{\sqrt{2}}\alpha_{R,r}$ and 
$\alpha_{A,r}=\frac{1}{\sqrt{2}}f_{2r}-\frac{1}{\sqrt{2}}\alpha_{R,r}$, where 
$f_{2r}$ is the independent vacuum noise. 
Here, $\alpha_D$ has a $1$-bit delay and $\alpha_A$ has a $-1$-bit delay. 
The optical delayline injects $\alpha_{A,r+1}$ and $\alpha_{D,r-1}$ into the $r$-th pulse $\alpha_{T,r}$. 
$\alpha_{F,r}=\frac{1}{\sqrt{2}}\alpha_{D,r-1}-\frac{1}{\sqrt{2}}\alpha_{A,r+1}$ is 
the mixed injection field from the delayline. 
The combined field at the injection beam splitter is represented by $\alpha'_r=\sqrt{1-R_B}\alpha_{T,r}+\sqrt{R_B}\alpha_{F,r}$. 
Here, $f_{1r}$ and $f_{2r}$ are vacuum fields in the Wigner representation, 
satisfying $\langle f_{ar}^* f_{br'}\rangle=\frac{1}{2}\delta_{ab}\delta_{r,r'}$. 
Next, we assume that the DOPO cavity round-trip time is $\Delta t$, and that the delay-line's reflectance is 
effectively simulated by a distributed coupling constant $J$, via $R_B=J\Delta t$. 
The traveling pulse with optical delay-line coupling model of Fig.\ref{model2} is then represented as
\begin{eqnarray}
\frac{d\alpha_r}{dt}|_C &=& -J\alpha_r+\frac{J}{2}(\alpha_{r-1}+\alpha_{r+1})+\frac{\sqrt{J}}{2}\xi_{C1,r}-\frac{\sqrt{J}}{4}\xi_{C1,r-1}\nonumber \\
&-&\frac{\sqrt{J}}{4}\xi_{C1,r+1}-\frac{\sqrt{J}}{4}\xi_{C2,r+1}+\frac{\sqrt{J}}{4}\xi_{C2,r-1}. 
\end{eqnarray}
Here, $\xi_{C1,r}$ and $\xi_{C2,r}$ come from the input noise of the BS and HBS respectively, 
and satisfy $\langle \xi_{Ca,r}(t)\xi_{Cb,r'}^*(t')\rangle=2\delta_{ab}\delta_{r,r'}\delta(t-t')$. 
We can see that this traveling-pulse model produces the same diffusion terms in the Fokker-Planck equation, as Eq.(\ref{wig1dnn}) does. 
We assume here that $R_B=J\Delta t \ll 1$.  

\end{document}